\documentclass[fleqn,10pt]{SelfArx} 
\usepackage[english]{babel}

\usepackage{subfigure}
\usepackage{url}
\usepackage{microtype}
\usepackage{pdflscape}
\usepackage{pdfpages}
\usepackage{rotating}
\usepackage{amsmath, amsthm, amssymb, amsfonts}

\newcommand{\beginsupplement}{%
        \setcounter{table}{0}
        \renewcommand{\thetable}{S\arabic{table}}%
        \setcounter{figure}{0}
        \renewcommand{\thefigure}{S\arabic{figure}}%
     }
\definecolor{color1}{RGB}{0,0,90} 
\definecolor{color2}{RGB}{0,20,20} 

\usepackage{hyperref} 
\hypersetup{hidelinks,colorlinks,breaklinks=true,urlcolor=blue,citecolor=color1,linkcolor=color1,bookmarksopen=false,pdftitle={Title},pdfauthor={Author}}
\usepackage{float}

\JournalInfo{Over-Optimization of Academic Publishing Metrics} 
\Archive{Draft Article}

\PaperTitle{Over-Optimization of Academic Publishing Metrics: Observing Goodhart's Law in Action} 
\Authors{Michael Fire\textsuperscript{1,2}* and Carlos Guestrin\textsuperscript{1}} 
\affiliation{\textsuperscript{1}\textit{Paul G. Allen School of Computer Science \& Engineering, University of Washington, Seattle, WA}} 
\affiliation{\textsuperscript{2}\textit{The eScience Institute, University of Washington, Seattle, WA}} 
\affiliation{*\textbf{Corresponding author}: fire@cs.washington.edu} 

\newcommand{\keywordname}{Keywords} 

\Abstract{The academic publishing world is changing significantly, with ever-growing numbers of publications each year and shifting publishing patterns. However, the metrics used to measure academic success, such as the number of publications, citation number, and impact factor, have not changed for decades.  Moreover, recent studies indicate that these metrics have become targets and follow Goodhart's Law, according to which ``when a measure becomes a target, it ceases to be a good measure.''  In this study, we analyzed over 120 million papers to examine how the academic publishing world has evolved over the last century. Our study shows that the validity of citation-based measures is being compromised and their usefulness is lessening. In particular, the number of publications has ceased to be a good metric as a result of longer author lists, shorter papers, and surging publication numbers. Citation-based metrics, such citation number and h-index, are likewise affected by the flood of papers, self-citations, and lengthy reference lists. Measures such as a journal's impact factor have also ceased to be good metrics due to the soaring numbers of papers that are published in top journals, particularly from the same pool of authors. Moreover, by analyzing properties of over 2600 research fields, we observed that citation-based metrics are not beneficial for comparing researchers in different fields, or even in the same department. Academic publishing has changed considerably; now we need to reconsider how we measure success.\\

\textbf{Multimedia Links} \\
\smallskip $\blacktriangleright$ \href{http://sciencedynamics.cs.washington.edu/}{\underline{Interactive Data Visualization}} \hspace{2mm}
 $\blacktriangleright$ \href{http://sciencedynamics.cs.washington.edu/code.html}{\underline{Code Tutorials}} \hspace{2mm}
 $\blacktriangleright$ \href{http://sciencedynamics.cs.washington.edu/fields_stat.html}{\underline{Fields-of-Study Features Table}}
 }

\begin{document}
\flushbottom 

\maketitle 

\tableofcontents 

\thispagestyle{empty} 

\section*{Introduction} 
\label{sec:intro}
\addcontentsline{toc}{section}{Introduction} 
\setcounter{section}{1}
In the last century, the academic publishing world has changed drastically in volume and velocity. The volume of papers has increased sharply from about 174,000 papers published in 1950 to over 7 million papers published 2014 (see Figure~\ref{fig:mag_papers}). Furthermore, the speed in which researchers can share and publish their studies has increased significantly. Today's researchers can publish not only in an ever-growing number of traditional venues, such as conferences and journals, but also in electronic preprint repositories and in mega-journals that provide rapid publication times~\cite{bjork2015have}.
\begin{landscape}
\begin{figure}[p]
  \centering
  \includegraphics[width=1.1\paperwidth]{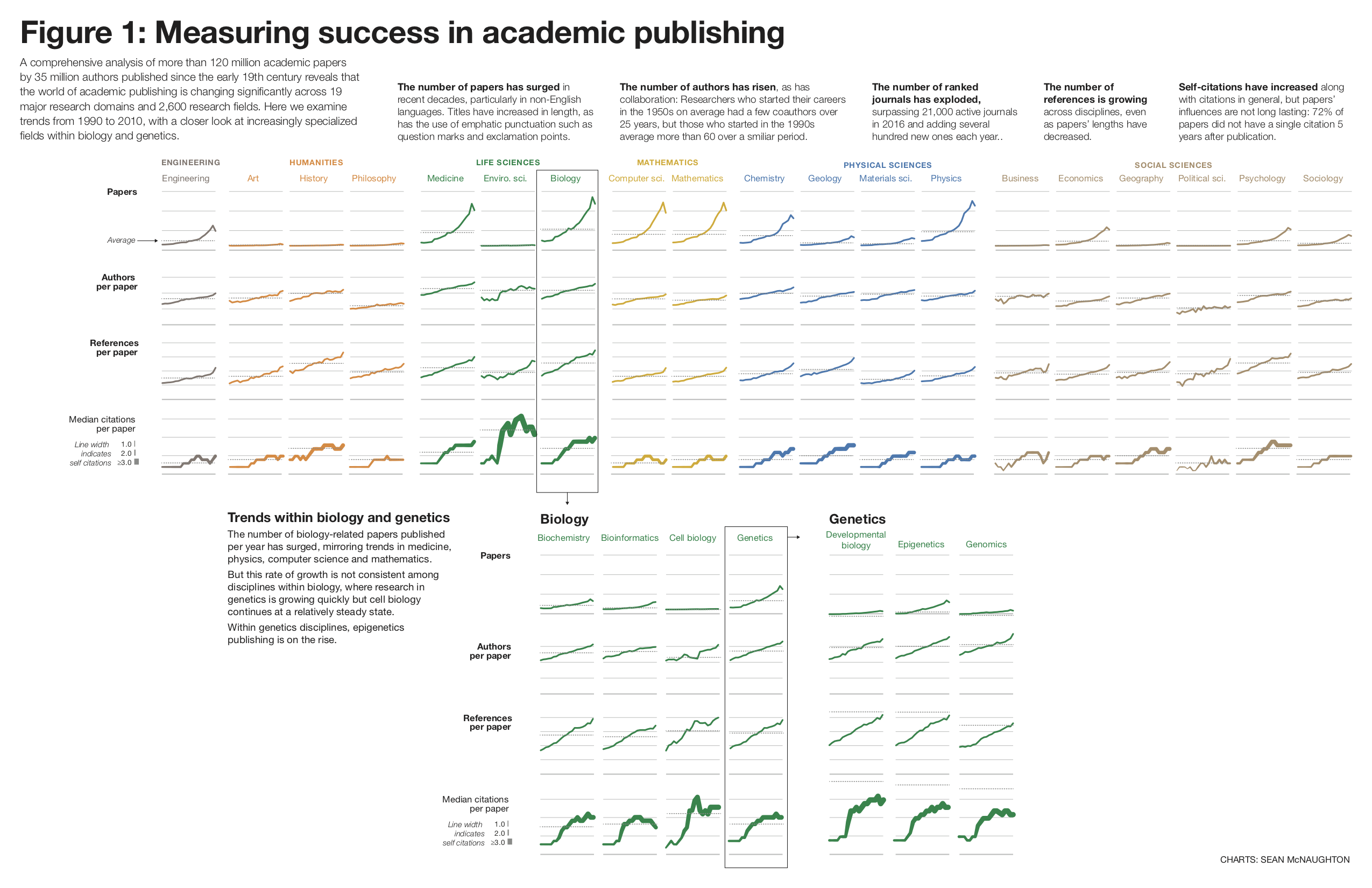}
\end{figure}
\end{landscape}

Along with the exponential increase in the quantity of published papers, the number of ranked scientific journals has increased to over 20,000 journals (see Figures~\ref{fig:journal_number} and~\ref{fig:jornual_number_mag}), and the number of published researchers has soared (see Figure~\ref{fig:author_number}). As part of this escalation, metrics such as the number of papers, number of citations, impact factor, h-index, and altmetrics are being used to compare the impact of papers, researchers, journals, and universities~\cite{roemer2012bibliometrics,hirsch2005index,garfield2005agony,wilsdon2016metric}. 
Using quantitative metrics to rank researchers contributes to a hypercompetitive research environment, which is changing academic culture - and not in a positive direction~\cite{edwards2017academic}.

Studies suggests that publication patterns have changed as a result of Goodhart's Law, according to which, ``When a measure becomes a target, it ceases to be a good measure''~\cite{edwards2017academic,biagioli2016watch}. Goodhart's Law, and its closely related Campbell's Law~\cite{campbell1979assessing}, influence many systems in our everyday life, including economical~\cite{mizen2003central}, educational~\cite{campbell1979assessing}, and other decision-making systems~\cite{chrystal2003goodhart}. As an example, Goodhart's Law can be found in the NYPD's manipulation of crime reports (the ``measure'') in order to improve crime statistics (the ``target'')~\cite{francescani2012nypd}.  Another example is found in the educational system, revealing that when ``test scores become the goal of the teaching process, they both lose their value as indicators of educational status and distort the educational process in undesirable ways''~\cite{campbell1979assessing}.  

Recent studies indicate that when measures become targets in academic publishing, the effectiveness of the measures can be compromised, and unwelcome and unethical behaviors may develop, 
such as salami publications~\cite{vsupak2013salami}, 
ghost authorships~\cite{schofferman2015ghost}, p-hacking~\cite{head2015extent}, metrics manipulation~\cite{bartneck2010detecting}, and even faking of peer reviews~\cite{haug2015peer}.

If the influence of Goodhart's Law on academia is indeed significant, then it should be possible to observe that academic entities, such as researchers and journals, will over-optimize their own measures to achieve a desired target. Similar to the consequences of making test scores a target, chasing after certain measures in the academic publishing world to desperately win the battle of ``\textit{impact or perish}''~\cite{biagioli2016watch} can have undesirable effects.

In this study, our main goal was to utilize new advances in data science tools to perform an in-depth and precise bottom-up analysis of academic publishing over the decades. Our comprehensive analysis ranged from micro to macro levels as we studied individual researchers' behaviors as well as behavioral changes within large research domains. Additionally, we wanted to uncover how and if Goodhart's Law has changed academic publishing. 

To achieve our research goals, we developed an open-source code framework to analyze data from several large-scale datasets containing over 120 million publications, with 528 million references and 35 million authors,\footnote{The number of authors was estimated according to the unique full names in the Microsoft Academic Graph dataset (see Section~\ref{sec:results_author_trends})} since the beginning of the 19\textsuperscript{th} century. This provided a precise and full picture of how the academic publishing world has evolved. In our analysis, we uncovered a wide variety of underlying changes in academia at different levels (see Figure 1): 

\begin{itemize}
  \item \textit{Papers} -- We observed that on average papers became shorter, yet other  features, such as titles, abstracts, and author lists, became longer (see Figures~\ref{fig:title_longer},~\ref{fig:paper_length},~\ref{fig:title_marks},~\ref{fig:paper_authors_alpha_order},~\ref{fig:paper_avg_abstract_len}, and~\ref{fig:paper_keywords}). Furthermore, both the number of references and the number of self-citations considerably increased (see Figures~\ref{fig:paper_avg_ref} and ~\ref{fig:paper_self_avg_max}). Moreover, the total number of papers without any citations at all increased sharply over time (see Figure~\ref{fig:no_citation_total}).
    
\item \textit{Authors} -- We noticed a sharp increase in the number of new authors. Moreover,  we observed a negative relation between researchers' career ages and the number of publications, i.e., early career researchers tended to publish more quickly than those later in their careers (see Figure~\ref{fig:author_pub_since_first}). Additionally, the average number of coauthors per author considerably increased over time (see Figure~\ref{fig:author_coauthors}). 

\item \textit{Journals} -- We observed a drastic increase in the number of ranked journals, with several hundreds of new ranked journals published each year (see Figures~\ref{fig:journal_number},~\ref{fig:jornual_number_mag}, and~\ref{fig:jornual_sjr_new_papers}). In addition, we observed that journal ranking changed significantly as average citations per document and SCImago Journal Rank (SJR) measures increased, while the h-index measure decreased over time. Moreover, the percentage of papers with returning authors increased sharply in recent years (see Figure~\ref{fig:top_journal_author_return}). For example, in Cell journal about 80\% of all papers published in 2016 included at least one author who had published in the journal before, while in 1980 this rate stood at less than 40\%.

\item \textit{Fields of Research} -- We analyzed the properties of 19 major research domains, such as art, biology, and computer science, as well as the properties of 2600 subdomains, such photography, genetic diversity, and computer animation. Our analysis revealed that different domains had widely ranging properties (see Figures~\ref{fig:field_number_papers} and~\ref{fig:field_num_papers_grid}), even within subdomains. For example, different subdomains of biology, and even subdomains of genetics, had surprisingly different average numbers of citations (see Figures~\ref{fig:field_l1_num_citation_grid}, and~\ref{fig:field_l2_num_citation_grid}).

\end{itemize}

These observations support the hypothesis that commonly used measures, such as the author's number of papers, h-index, and citation number, have become targets. Researchers can increase their average number of papers by writing shorter papers with more coauthors. The h-index can be boosted by increasing the number of papers and self-citations. A higher citation number is attained with longer titles and abstracts, as well as more references (see Section~\ref{sec:paper_results} and Figure~\ref{fig:paper_corr}). 

It is time to consider how we judge academic papers. Citation-based measures have been the standard for decades, but these measures are far from perfect. In fact, our study shows that the validity of citation-based measures is being compromised and their usefulness is lessening. Goodhart's Law is in action in the academic publishing world.

The remainder of the paper is organized as follows: In Section~\ref{sec:related}, we provide an overview of related studies. In Section~\ref{sec:methods}, we describe the datasets, methods, algorithms, and experiments used throughout this study. Next, in Section~\ref{sec:results}, we present the results of our study. Afterwards, in Section~\ref{sec:diss}, we discuss the obtained results. Lastly, in Section~\ref{sec:conclusions}, we present our conclusions. 


\section{Related Work}
\label{sec:related}
This research is a large-scale scientometrics study (also referred to as the ``science of science''~\cite{fortunato2018science}). Scientometrics is the study of quantitative features and characteristics of scientific research. In this section, we give a short overview of the relevant scientometric papers to this study. We present studies that analyze changes in academic publications in recent years (see Section~\ref{sec:related_trends}), and we provide an overview of common metrics that measure the impact of published papers (see Section~\ref{sec:related_metrics}).

\subsection{Changes in Publication Trends}
\label{sec:related_trends}
One prevalent and increasing trend is to publish papers in preprint repositories, such as arXiv, bioRxiv, and SSRN. For example, the use of arXiv surged from 4,275 papers in September 2006 to 10,570 papers in September 2017~\cite{arXiv2018}. Another common trend is to publish papers in mega-journals, such as PLOS ONE and Nature's Scientific Reports. Mega-journals are a new type of scientific journal that publishes peer-reviewed, open-access articles, where the articles have been reviewed for scientific trustworthiness, but not for scientific merit. Mega-journals accelerate review and publication times to 3-5 months and usually have high acceptance rates of over 50\%~\cite{bjork2015have}. In the first quarter of 2017, over 11,000 papers were published in PLOS ONE and Scientific Reports~\cite{davis2017}.  

Another observable trend is that more and more papers are written by hundreds of authors. The recent Laser Interferometer Gravitational-Wave Observatory (LIGO) paper~\cite{abbott2016observation} has over 1000 authors~\cite{castelvecchiligo}. Robert Aboukhalil measured this trend ~\cite{aboukhalil2014rising} and discovered that the average number of authors of academic papers has increased sharply since the beginning of the 20\textsuperscript{th} century. While papers' average number of authors has gone up over time, not all the authors have significantly contributed to the paper. Honorary and ghost authors are prevalent. Wislar et al. found such evidence in biomedical journals~\cite{wislar2011honorary}, and similar findings were observed by Kennedy et al.~\cite{kennedy2014honorary} and by Vera-Badillo et al.~\cite{vera2016honorary}. The Economist recently published an article titled ``Why research papers have so many authors''~\cite{economist2016}.  

Lewison and Hartley~\cite{lewison2005s} analyzed how papers' titles have changed over time. They discovered that titles' lengths have been increasing, along with the percentage of titles containing colons. Additionally, Gwilym Lockwood observed that ``articles with positively-framed titles, interesting phrasing, and no wordplay get more attention online''~\cite{lockwood2016academic}. 

Additionally, many studies have focused on how publication trends have changed over time, often  focusing on specific geographical areas, various demographic characteristics, specific research domains, or specific journals. For example, G\'{a}lvez et al.~\cite{galvez2000scientific} utilized the Science Citation Index to understand publication patterns in the developing world. Jagsi et al.~\cite{jagsi2006gender} studied the gender gap in authorship of academic medical literature over 35 years. They discovered that the percentage of first and last authors who were women increased from 5.9\% and 3.7\% in 1970 to 29.3\% and 19.3\%, respectively, in 2004. Johnson et al.~\cite{johnson2016publication} studied publication trends in top-tier journals of higher education.  Peter Aldhous analyzed publications in the National Academy of Sciences (PNAS) journal, to consider the influence of an``old boys' club'' mentality~\cite{aldhous2014}.

Our study is greatly influenced by a recent study by Edwards and Roy~\cite{edwards2017academic}, who observed that academia has become a hypercompetitive environment that can lead to unethical behaviors. The driving force behind such behaviors is to manipulate the metrics that measure the research's impact solely to increase the quantitative measures (and hence the status) of the research.

\subsection{Success Metrics and Citation Trends}
\label{sec:related_metrics}
Over the years, various metrics have been proposed to measure papers, journal importance, and authors' impact. One of the most straightforward and commonly utilized measure is to simply count the researcher's number of publications. Another common metric is the citation number, either of a particular paper or the total citations received by all the author's papers. However, not all citations are equal~\cite{yan2010weighted}. Moreover, different research fields have different citation metrics, and therefore comparing them creates a problem: ``The purpose of comparing citation records is to discriminate between scientists''~\cite{lehmann2006measures}.  

One of the best-known and most-used measures to evaluate journals' importance is the impact factor, devised over 60 years ago by Eugene Garfield~\cite{garfield2005agony}. The impact factor measures the frequency in which an average article in a journal has been cited in a specific year. Over time, the measure has been used to ``evaluate institutions, scientific research, entire journals, and individual articles''~\cite{garfield2003meaning}.  Another common metric to measure a researcher's output or a journal's impact is the h-index, which measures an author's or a journal's number of papers that have at least h citations each~\cite{hirsch2005index}.  It has been shown that the h-index can predict academic achievements~\cite{hirsch2007does}. 

The above measures have been the standard for measuring academic publishing success. According to recent studies, and following Goodhart's Law, these metrics have now become targets, ripe for manipulation~\cite{edwards2017academic, biagioli2016watch,fong2017authorship}.  All types of manipulative methods are used, such as increasing the number of self-citations~\cite{bartneck2010detecting}, increasing the number of publications by slicing studies into the smallest measurable quantum acceptable for publication~\cite{nature2005cost}, indexing false papers~\cite{delgado2014g}, and merging papers on Google Scholar~\cite{van2016h}. Indeed, a recent study by Fong and Wilhite~\cite{fong2017authorship}, which utilized data from over 12,000 responses to a series of surveys sent to more than 110,000 scholars from eighteen different disciplines, discovered ``widespread misattribution in publications and in research proposals.'' Fong and Wilhite's findings revealed that the majority of researchers disapprove of this type of metric manipulation, yet many feel pressured to participate; other researchers blandly state ``that it is just the way the game is played''~\cite{fong2017authorship}.

Due to many common metric shortcomings, various alternative measures have been proposed. For example, the q-index~\cite{bartneck2010detecting} and w-index~\cite{wu2010w} are alternatives to the h-index. Likewise, the SJR indicator~\cite{falagas2008comparison} and simple citation distributions~\cite{lariviere2016simple} are offered as alternatives to the impact factor. Senior employees at several leading science publishers called upon journals to restrain from using the impact factor and suggested replacing it with simple citation distributions~\cite{lariviere2016simple,callaway2016}. Similarly, the altmetric\footnote{\url{https://www.altmetric.com/}} was proposed as an alternative metric to the impact factor and h-index. The altmetric~\cite{griffin2017altmetrics} is a generalization of article-level metrics and considers other aspects of the impact of the work, such as the number of downloads, article views, mentions in social media, and more. The altmetric measure has gained in popularity in recent years, and several large publishers have started providing this metric to their readers.  Additionally, Semantic Scholar\footnote{\url{https://www.semanticscholar.org}} offers various measures to judge papers and researchers' influence. A thorough report regarding potential uses and limitations of metrics was written by Wilsdon et al.~\cite{wilsdon2016metric}. Additionally, an overview of the changing scholarly landscape can be found in Roemer and Borchardt's study~\cite{roemer2012bibliometrics}.

Even with their many known shortcomings~\cite{wilsdon2016metric,lehmann2006measures,seglen1997impact,byrne2017,Hecht1998TheJ}, measures such as the impact factor, citation number, and h-index are still widely used. For example, the Journal Citation Reports publishes annual rankings based on journals' impact factors, and it continues to be widely followed.

\section{Methods and Experiments}
\label{sec:methods}
\subsection{Datasets}
\label{sec:datasets}
\subsubsection{The Microsoft Academic Graph (MAG) Dataset }
\label{sec:MAG}

In this study we primarily utilized the Microsoft Academic Graph (MAG)~\cite{Sinha2015AnOO}, which was released as part of the 2016 KDD Cup~\cite{kdd2016}. The large-scale MAG dataset contains scientific publication records of over 120 million papers, along with citation relationships among those publications as well as relationships among authors, institutions, journals, conferences, and fields of study. In addition, the MAG dataset contains every author's sequence number for each paper's authors list. Furthermore, the dataset contains field-of-study hierarchy ranking with four levels, L0 to L3, where L0 is the highest level, such as a research field of computer science, and L3 is the lowest level, such as a research field of decision tree~\cite{kdd2016}.

Even though the MAG dataset contains papers that were published through 2016, we wanted to use years in which the data was the most comprehensive, so we focused our analysis on 120.7 million papers which were published through the end of 2014. Furthermore, we noted that the dataset contains many papers that are news items, response letters, comments, etc. Even though these items are important, they can affect a correct understanding of the underlying trends in scientific publications. Therefore, we focused our research on a dataset subset, which consists of over 22 million papers. This subset contains only papers which have a Digital Object Identifier (DOI) and at least 5 references. Additionally, while calculating various authors' properties, we primarily considered only the 22.4 million authors with unique author ID values in the selected papers’ subset.

\subsubsection{The AMiner Dataset}
\label{sec:AMIner}
The AMiner open academic graph dataset~\cite{Tang2008ArnetMinerEA} contains data from over 154 million papers. The dataset contains various papers' attributes, such as titles, keywords, abstracts, venues, languages, and ISSNs. In our study, we primarily utilized the AMiner dataset to analyze papers' abstracts, to estimate papers' lengths, and to compare results with those obtained using the MAG dataset in order to validate the existence of observed patterns in both datasets.

\subsubsection{The SCImago Journal Rank Dataset}
\label{sec:SJR}
To better understand trends in journal publications, we used the SCImago Journal Ranking (SJR) open dataset~\cite{Butler2008FreeJT}.\footnote{\url{https://www.scimagojr.com/journalrank.php}}  This dataset contains details of over 23,000 journals with unique names between 1999 and 2016. For each journal, the SJR dataset contains the journal's SJR value, the number of published papers, the h-index, and the number of citations in each year. Additionally, the SJR dataset contains the best quartile (ranked from Q1 to Q4) of each journal. The quartile rank is typically used to compare and rank journals within a given subject category.

\subsubsection{The Join Dataset}
\label{sec:join}
To match the MAG journal IDs with their correlated various ranking measures, such as h-index and SJR, we joined all three datasets in the following manner: First, we joined the MAG and AMiner datasets by matching unique DOI values. Then, we matched ISSN values between the MAG-AMiner joined dataset with the SJR dataset.

\subsection{Code Framework}
\label{sec:framework}
To analyze the above MAG and AMiner large-scale datasets, we developed an open source framework written in Python, which provided an easy way to query the datasets. The framework utilizes TuriCreate's SFrame dataframe objects~\cite{Low2010GraphLabAN} to perform big-data analysis on tens of millions of records to calculate how various properties have changed over time. For example, we used SFrame objects to analyze how the average number of authors and title lengths evolved. However, while SFrame is exceptionally useful for calculating various statistics using all-papers features, it is less convenient and less computationally cost effective for performing more complicated queries, such as calculating the average age of the last authors in a certain journal in a specific year. 

To perform more complex calculations, we loaded the datasets into the MongoDB database.\footnote{\url{http://www.mongodb.com}}  Next, we developed a code framework that easily let us obtain information on papers, authors, paper collections, venues, and research fields. The framework supports calculating complex features of the above object in a straightforward manner. For example, with only a few and relative simple lines of Python code, we were able to calculate the average number of coauthors per author in a specific year for authors who started their career in a specific decade. An overview of our code framework is presented in Figure~\ref{fig:framework}. 

To make our framework accessible to other researchers and to make this study completely reproducible, we have written Jupyter Notebook tutorials which demonstrate how the SFrame and MongoDB collections were constructed from the MAG, AMiner, and SJR datasets.  

\subsection{Analysis of Publication Trends}
\label{sec:pub_trends}
We used our developed code framework to explore how papers, authors, journals, and research fields have evolved over time. In the following subsections, we describe the specific calculations that were performed. Moreover, our Supplementary Materials section includes the precise code implementations which were used to obtain most of our results and to create the figures presented throughout this study.

\subsubsection{Paper Trends}
\label{sec:paper_trends}
To explore how the quantity and structure of academic papers have changed over time, we performed the following: First, we calculated how many papers were published in the MAG dataset every year. Then, we detected the language of each paper's title and calculated the number of papers in each language. Next, we calculated the following paper features over time:

\begin{itemize}
    \item Average number of words in titles and average number of characters per word (for papers with English titles)
    
    \item Percentage of titles that used question or exclamation marks (for papers with English titles)
    
    \item Average number of authors
    
    \item Percentage of papers in which authors appear in alphabetical order
    
    \item Average number of words in abstracts
    
    \item Average number of keywords
    
    \item Average number of references
    
    \item Length of papers
\end{itemize}
In addition, we utilized the papers with existing field-of-research values, matching the papers to their corresponding fields in order to identify each paper's top level (L0) research field. Using the top-level data, we were able to estimate the number of multidisciplinary papers that had more than one L0 research field. Afterwards, we calculated the percentage and total number of papers with no citations after 5 years, as well as the overall percentage of papers with self-citations over time.\footnote{We define paper A as self-citing paper B if at least one of the authors of A is also an author of B.}  Lastly, to better understand how citation patterns have changed across generations, we calculated the citation distributions after 10 years for each decade between 1950 and 2000.   

Additionally, we selected all the papers in the Join dataset that had valid features\footnote{We selected only papers having English titles and abstracts, existing author lists, references, and valid lengths. Additionally, we checked if the paper's title contained question or exclamation marks.} and were published between 1990 and 2009. Using the selected papers, we calculated the Spearman correlations among the title lengths, author numbers, reference numbers, overall lengths, and number of citations after 5 years. 
The results of the above described calculations are presented in Section~\ref{sec:paper_results}. Moreover, the code implementation is provided in the ``Part III - A: Analyzing Changing Trends in Academia – Paper Trends'' Jupyter Notebook (see Section~\ref{sec:code}).

\subsubsection{Author Trends}
\label{sec:authors_trends}
To study how authors' behaviors and characteristics have changed, we performed the following: First, we calculated how the number of new authors has changed over time. Second, for all authors who published their first paper after 1950, we divided the authors into groups according to each author's academic birth decade, i.e., the decade in which an author published his or her first paper. Next, for each group of authors with the same academic birth decade, we analyzed the following features: 

\begin{itemize}
    \item Average number of papers the authors in each group published $n$ years after they began their careers, for $\forall n \in[0,30]$. We performed these group calculations taking into account all papers, as well as only papers with at least 5 references.  
    
    \item 	Average number of conference and journal papers each group published $n$ years after they began their careers, for $\forall n \in[0,30]$
    
    \item 	Average number of coauthors each group had $n$ years after they began their careers, for $\forall n \in[0,30]$
    
    \item Authors' median sequence number each group had $n$ years after they began their careers, for $\forall n \in[0,60]$. Additionally, we calculated the average percentage of times the authors in each group were first authors.
\end{itemize}

The results of the above described calculations are presented in Section~\ref{sec:results_author_trends}. Moreover, the code implementation is provided in the ``Part III - B: Analyzing Changing Trends in Academia - Author Trends'' Jupyter Notebook (see Section~\ref{sec:code}).

\subsubsection{Journal Trends}
\label{sec:journal_trends}
To investigate how journal publication trends have changed over time, we used the SJR dataset to calculate the following features between 1999 and 2016:
\begin{itemize}
    \item Number of journals with unique journal IDs that were active in each year 
    \item Number of new journals that were published each year
    \item Average and maximal number of papers in each journal
\end{itemize}

Additionally, we utilized the SJR dataset to calculate how the journals' best quartile, average h-index, average SJR, and average citation number ($\frac{Citation Number}{Documents Number}\mbox{  (2 years)}$) metrics changed between 1999 and 2016. 

Furthermore, we selected the 40 journals with the highest SJR values in 2016 and matched them to their corresponding journal IDs in the MAG dataset by matching each journal’s ISSN and exact name in the MAG-AMiner joined dataset.\footnote{The top journal name was compared to the journal’s name in the MAG dataset.} Then, for the matching journal IDs, we calculated the following features over time, for all papers that were published in the selected top journals:
\begin{itemize}
    \item First and last authors' average career age 
    \item Percentage of papers in which the first author had previously published in the one of the top journals
    \item Percentage of papers in which the last author had previously published in the one of the top journals
\end{itemize}

The results of the above described calculations are presented in Section~\ref{sec:results_journal_trends}. Moreover, the code implementation is provided in the ``Part III - C: Analyzing Changing Trends in Academia - Journal Trends'' Jupyter Notebook (see Section~\ref{sec:code}).

Additionally, for over 8,400 journals with at least 100 published papers with 5 references, we calculated the following features over time:
\begin{itemize}
    \item Number of papers
    \item Number of authors
    \item Top keywords in a specific year
    \item First/last/all authors average or median academic age
    \item Average length of papers
    \item Percentage of returning first/last/all authors, i.e., those who had published at least one prior paper in the journal
\end{itemize}

We developed a \href{http://sciencedynamics.cs.washington.edu/}{website} with an interactive interface, which visualizes how the above features changed for each journal (see Section~\ref{sec:code}).

\subsubsection{Field-of-Research Trends}
\label{sec:fields_trends}
We utilized the MAG dataset field-of-study values and the hierarchical relationship between various fields to match papers to their research fields in various levels (L0-L3). Then, for each field of study in its highest hierarchical level (L0), we calculated the following features over time: number of papers, number of authors, number of references, and average number of citations after five years. Next, we focused on the field of biology, which is in the L0 level.  For all the L1 subfields of biology, we repeated the same feature calculations as in the previous step. Afterwards, we focused on genetics. For all the L2 subfields of genetics, we repeated the same feature calculations as in the previous step.

Additionally, to better understand the differences in citation patterns of various fields of research, we performed the following: For each field of study with at least 100 papers published in 2009, we calculated the following features using only papers that were published in 2009 and had at least 5 references: 
\begin{itemize}
    \item Number of papers 
    \item Number of authors
    \item Median and average number of citations after 5 years  
    \item Maximal number of citations after 5 years
\end{itemize}
The full features of over 2600 L3 fields of study are presented in Table~\ref{tab:l3}.

The results of the above described calculations are presented in Section~\ref{sec:results_fields_trends}. Moreover, the code implementation is provided in the ``Part III - D: Analyzing Changing Trends in Academia - Research Fields'' Jupyter Notebook (see Section~\ref{sec:code}).

\section{Results}
\label{sec:results}
In the following subsections, we present all the results for the experiments which were described in Section~\ref{sec:pub_trends}. Additional results are presented in the Supplementary Materials.

\subsection{Results of Paper Trends}
\label{sec:paper_results}
\begin{figure}
\centering 
\includegraphics[width=\linewidth]{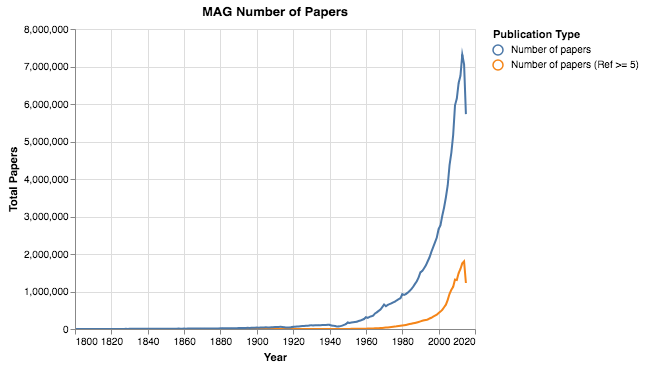}
\caption{\textbf{The Number of Papers over Time.} The total number of papers has surged exponentially over the years.}
\label{fig:mag_papers}
\end{figure}

\begin{figure}
\centering 
\includegraphics[width=\linewidth]{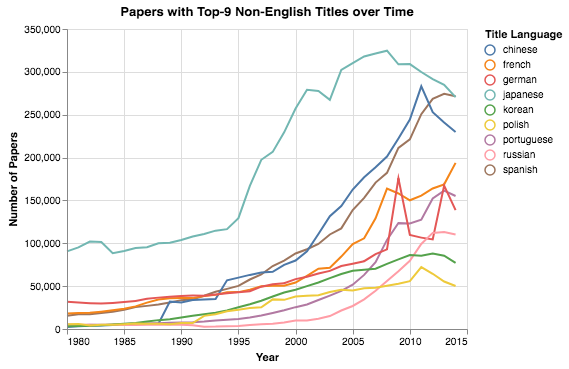}
\caption{\textbf{Papers with Top-9 Non-English Titles.} Increasingly, more papers have non-English titles. }
\label{fig:non_english_papers}
\end{figure}

In recent years there has been a surge in the number of published academic papers, with over 7 million new papers each year and over 1.8 million papers with at least 5 references (see Figure~\ref{fig:mag_papers}).\footnote{There is a decline in the number of papers after 2014, probably due to missing papers in the MAG dataset, which was released in 2016.}  Additionally, by analyzing the language of the papers' titles, we observed a growth in papers with non-English titles (see Figure~\ref{fig:non_english_papers}). 

\begin{figure*}
\centering
\subfigure{%

\includegraphics[width=0.45\linewidth]{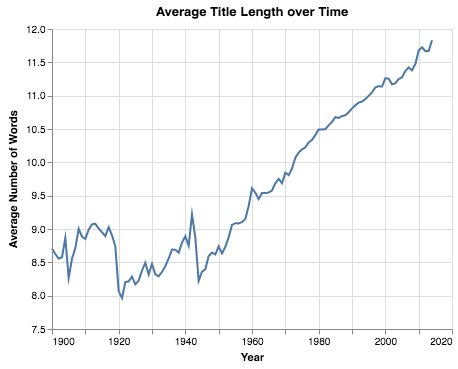}}%
\qquad
\subfigure{%
\includegraphics[width=0.45\linewidth]{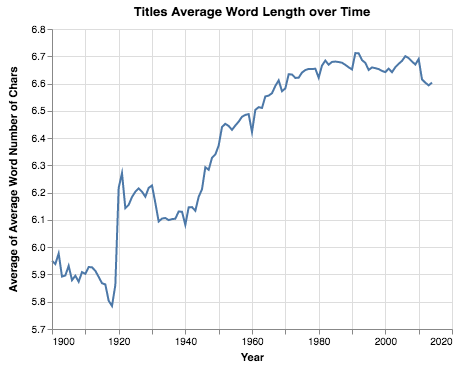}}%
\caption{\textbf{Average Title Length over Time.} A paper's average title length increased from 8.71 words to over 11.83 words. Moreover, the average word length increased from 5.95 characters to 6.6 characters per title word.}
\label{fig:title_longer}
\end{figure*}

As described in Section~\ref{sec:paper_trends}, we analyzed how various properties of academic papers have changed over time to better understand how papers' structures have evolved. In this analysis, we discovered that papers' titles became longer, from an average of 8.71 words in 1900 to an average of 11.83 words in 2014 (see Figure~\ref{fig:title_longer}). Moreover, the average number of characters per word increased from 5.95 characters per average title word in 1900 to 6.6 characters per average title word in 2014 (see Figure~\ref{fig:title_longer}). Additionally, we observed that in recent years the percentage of papers with question or exclamation marks increased sharply, from less than 1\% of all papers in 1950 to over 3\% of all papers in 2013 (see Figure~\ref{fig:title_marks}). Furthermore, the usage of interrobangs (represented by ?! or !?) also increased sharply, from 0.0005\% in 1950 to 0.0037\% in 2013 (see  Figure~\ref{fig:title_marks}).

\begin{figure} 
\centering 
\includegraphics[width=\linewidth]{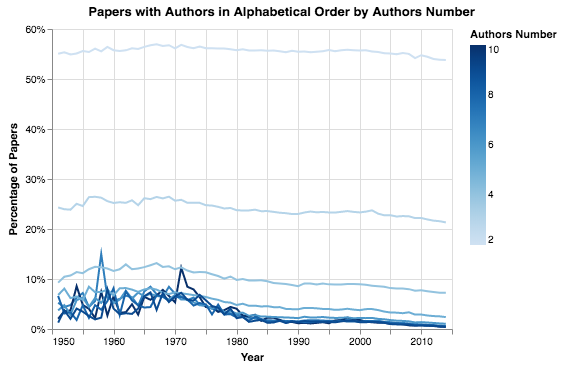}
\caption{\textbf{Percentage of Papers with Author Lists in Alphabetical Order, Grouped by the Number of Authors.} The higher the number of authors, the less likely the authors will be organized alphabetically.  }
\label{fig:paper_authors_alpha_order_by_authors}
\end{figure}

We explored how the number and order of the authors list has changed over time. The number of authors for papers with at least 5 references more than tripled over the years, from an average of 1.41 authors to an average of 4.51 authors per paper between 1900 and 2014, respectively (see Figure ~\ref{fig:paper_authors_number}). Also, the maximal number of authors for a single paper in each year increased sharply over time, especially in recent years (see Figure~\ref{fig:paper_max_authors}). In fact, some recent papers actually listed over 3000 authors. Moreover, we observed that the percentage of author lists ordered alphabetically decreased in recent years, from 43.5\% of all papers published in 1950 to 21\% of all papers published in 2014 (see Figure~\ref{fig:paper_authors_alpha_order}). Furthermore, we discovered that with a higher number of authors, it is less likely that the authors list will be ordered alphabetically (see Figure~\ref{fig:paper_authors_alpha_order_by_authors}). For example, in 2014 only about 1\% of papers with six authors were ordered alphabetically.

\begin{figure} [ht]
\centering 
\includegraphics[width=\linewidth]{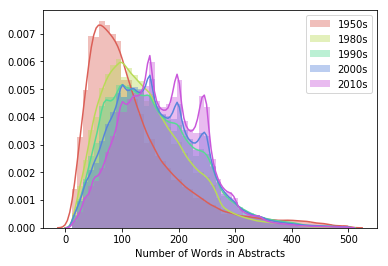}
\caption{\textbf{Distribution over Time of the Number of Words in Abstracts.} Over time, papers' abstracts have tended to become longer. }
\label{fig:paper_abstract_len_dists}
\end{figure}

When calculating how the abstracts of papers have changed over time, we discovered that the abstract length increased from an average of 116.3 words in 1970 to an average of 179.8 words in 2014 (see Figure~\ref{fig:paper_avg_abstract_len}). Moreover, with each decade since 1950, the distributions shifted to the right, showing that papers with longer abstracts of 400 and even 500 words have become more common over time (see Figure~\ref{fig:paper_abstract_len_dists}). Additionally, we analyzed how the number of keywords in papers has changed. We discovered that both the number of papers containing keywords increased, as well as the average number of keywords per paper (see Figure~\ref{fig:paper_keywords}).

\begin{figure*}%
\centering
\subfigure{%

\includegraphics[width=0.45\linewidth]{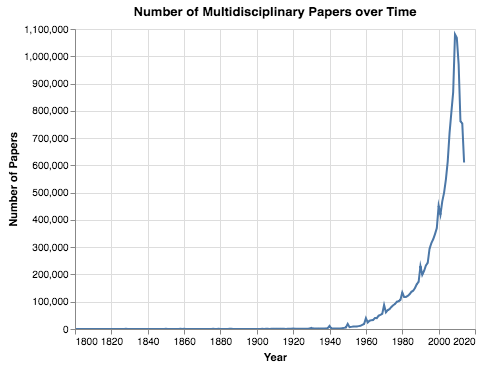}}%
\qquad
\subfigure{%
\includegraphics[width=0.45\linewidth]{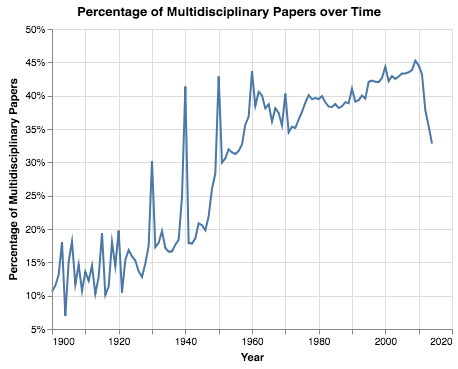}}%
\caption{\textbf{The Number and Percentage of Multidisciplinary Papers over Time.} Between 1900 and 2010, both the number and percentage of multidisciplinary papers increased over time. }
\label{fig:paper_multidis}
\end{figure*}

By estimating the percentage and number of multidisciplinary papers over time, we discovered an increase in the number of multidisciplinary papers until 2010, followed by a sharp decrease  (see Figures~\ref{fig:paper_multidis} and~\ref{fig:paper_multidis_l0_l1}). After performing further analysis, we believe the decline in the number of multidisciplinary papers is a result of papers with missing keywords in the MAG dataset, such as papers that were published in the PLOS ONE journal. These papers have dynamically changing keywords in the online version, but not in the offline version.

By examining how the number of references has changed over time, we observed a sharp increase in the average number of references per paper (see Figure~\ref{fig:paper_avg_ref}). In addition, by analyzing the reference number distributions grouped by publishing decade, we can observe that higher numbers of references have become increasingly common. For example, in 1960 few papers had over 20 references, but by 2010 many papers had over 20 references, and some over 40 references (see Figure~\ref{fig:paper_ref_dists}). 

\begin{figure*}%
\centering
\subfigure{%

\includegraphics[width=0.45\linewidth]{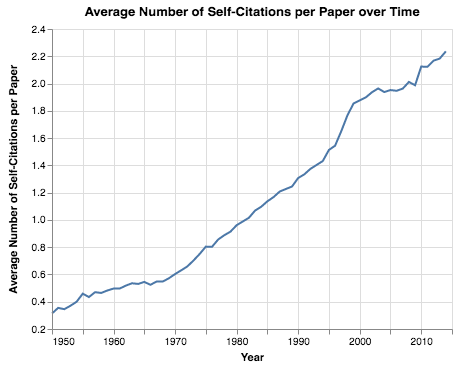}}%
\qquad
\subfigure{%
\includegraphics[width=0.45\linewidth]{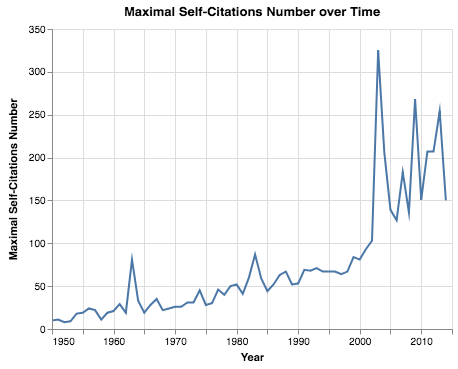}}%
\caption{\textbf{The Average and Maximal Number of Self-Citations.} Both the average and maximal number of self-citations increased over time. }
\label{fig:paper_self_avg_max}
\end{figure*}

We also examined how self-citation trends have changed, and we observed that both the total number of self-citations and the percentage of papers with self-citations increased significantly (see Figure~\ref{fig:paper_self_total_percentage}). Also, the average number of self-citations per paper, as well as the maximal number of self-citations in each year, increased sharply (see Figure~\ref{fig:paper_self_avg_max}). For example, about 3.67\% of all papers in 1950 contained at least one self-citation, while 8.29\% contained self-citations in 2014 (see Figure~\ref{fig:paper_self_total_percentage}). Moreover, the maximal number of self-citations in a single paper increased sharply from 10 self-citations in a paper published in 1950 to over 250 self-citations in a paper published in 2013 (see Figure~\ref{fig:paper_self_avg_max}).

\begin{figure*}%
\centering
\subfigure{%

\includegraphics[width=0.45\linewidth]{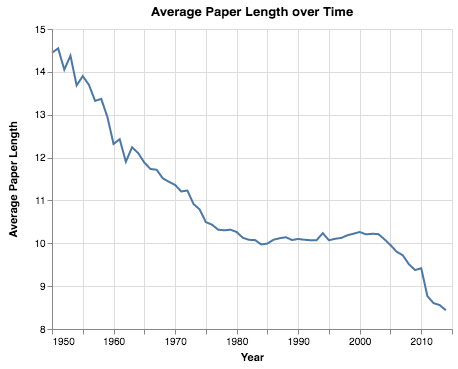}}%
\qquad
\subfigure{%
\includegraphics[width=0.45\linewidth]{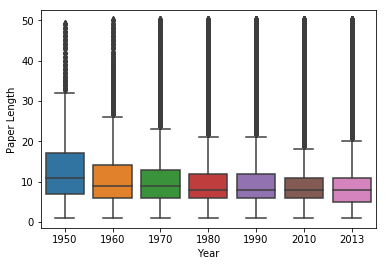}}%
\caption{\textbf{Papers' Lengths.} Both the papers' average and median lengths decreased over time. }
\label{fig:paper_length}
\end{figure*}

By using the AMiner dataset to analyze how papers' lengths have changed, we discovered that the average and median length of papers decreased over time (see Figure~\ref{fig:paper_length}). The average length of a paper was 14.4, 10.1, and 8.4 pages in 1950, 1990, and 2014, respectively.

\begin{figure}
\centering 
\includegraphics[width=\linewidth]{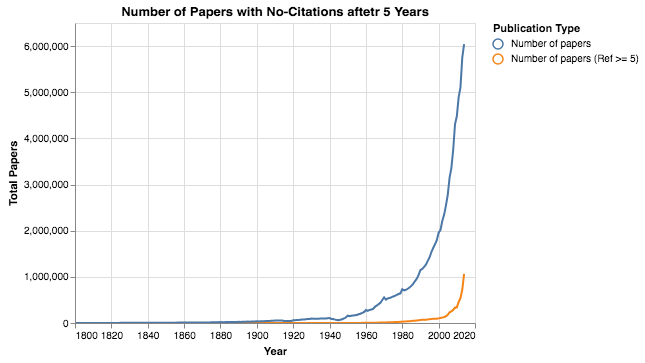}
\caption{\textbf{Total Number of Papers with No Citations after 5 Years.} The number of papers with increased sharply over time. }
\label{fig:no_citation_total}
\end{figure}

By analyzing citation patterns over time, we discovered that the percentage of papers with no citations after 5 years decreased (see Figure~\ref{fig:no_citations}). Nevertheless, still about 72.1\% of all papers published in 2009, and 25.6\% of those with at least 5 references, were without any citations after 5 years (see Figure~\ref{fig:no_citations}). Moreover, the total number of papers without any citations increased sharply (see Figure~\ref{fig:no_citation_total}).

\begin{figure}
\centering 
\includegraphics[width=\linewidth]{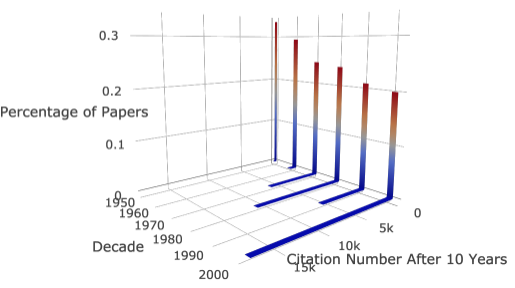}
\caption{\textbf{Citation Distributions over Time.} The citation distributions of different decades show notable changes.  }
\label{fig:citations_3d}
\end{figure}

Additionally, by analyzing the citation distributions of papers published in different decades, we discovered citation distributions changed notably over time (see Figure~\ref{fig:citations_3d}).

Lastly, using the properties of over 3.29 million papers published between 1950 and 2009, we discovered positive correlations among the papers' citation numbers after 5 years and the following features: (a) title lengths ($\tau_s=0.1$); (b) author numbers ($\tau_s=0.22$); (c) abstract lengths ($\tau_s=0.26$); (d) keyword numbers ($\tau_s=0.15$); (e) reference numbers ($\tau_s=0.48$); (e) paper lengths ($\tau_s=0.13$); and (f) use of question or exclamation marks ($\tau_s=0.022$) (see Figure~\ref{fig:paper_corr}).\footnote{Similar correlation values were obtained by calculating the correlations for papers published in a specific year.}

\subsection{Results of Author Trends}
\label{sec:results_author_trends}
\begin{figure}
\centering 
\includegraphics[width=\linewidth]{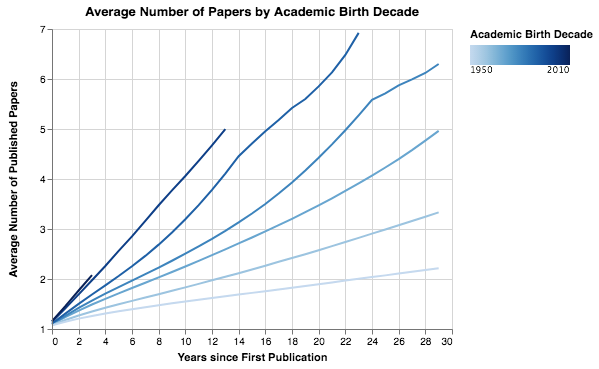}
\caption{\textbf{Average Number of Papers by Authors' Academic Birth Decades.} With each decade, the rate of paper publication has increased.   }
\label{fig:author_pub_since_first}
\end{figure}

By analyzing the number of new authors each year, we discovered a sharp increase over time, with several million new authors publishing each year in recent years (see Figure~\ref{fig:author_number}).\footnote{It is possible that the same author has several MAG author IDs.} Additionally, when analyzing the trends grouped by the authors' academic birth decades, we discovered a significant increase in the average number of published papers for the younger birth decades (see Figure~\ref{fig:author_pub_since_first}). For example, researchers who started their careers in 1950 published on average 1.55 papers in a time period of 10 years, while researchers who started their careers in 2000 published on average 4.05 papers in the same time frame. Furthermore, we observed that authors who started their careers after 1990 tended to publish more in conferences in the first years of their career than their more senior peers who started their careers in the 1950s or 1970s (see Figure~\ref{fig:author_journal_conf}). For example, researchers who started their careers in the 1970s published on average about 2 conference papers and 1.65 journal papers after 10 years; researchers who started their careers in the 2000s published about 4 conference papers and 2.59 journal papers in the same time frame.

\begin{figure}
\centering 
\includegraphics[width=\linewidth]{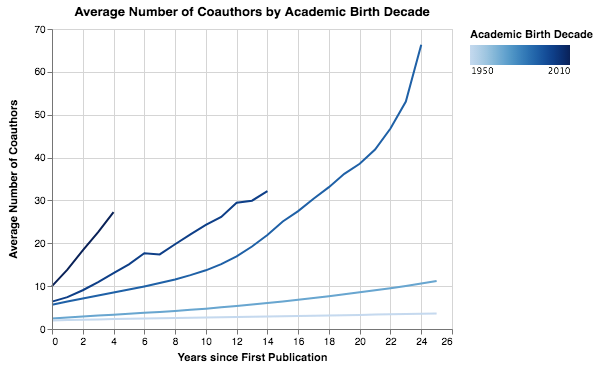}
\caption{\textbf{Average Number of Coauthors by Academic Birth Decade.} The average number of coauthors has considerably increased over the decades.   }
\label{fig:author_coauthors}
\end{figure}

We can also observe that the average number of coauthors has considerably increased over the decades (see Figure~\ref{fig:author_coauthors}). Moreover, we can notice that researchers who started their careers in the 1950s and 1970s had on average only few coauthors over a period of 25 years, while researchers who started their careers in the 1990s had over 60 coauthors in the same career length of 25 years (see Figure~\ref{fig:author_coauthors}).

\begin{figure} 
\centering 
\includegraphics[width=\linewidth]{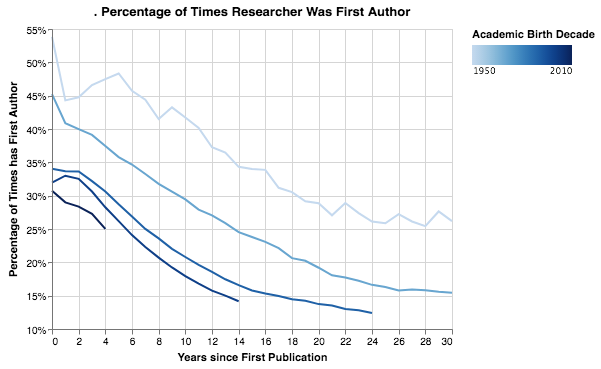}
\caption{\textbf{Percentage of Times Researcher was First Author.} We can observe that over time on average the percentage of senior researchers as first authors declined. Moreover, in the same time intervals, the percentage of times recent generations of researchers were first authors declined compared to older generations.  }
\label{fig:author_first_author_percentage}
\end{figure}

Lastly, by exploring how author sequence numbers evolved, we discovered that with seniority, the researchers' median sequence number increased (see Figure~\ref{fig:author_med_seq}). Additionally, with seniority, the percentage of published papers with the researcher listed as the first author decreased (see Figure~\ref{fig:author_first_author_percentage}). Moreover, by looking at the decade researchers started their careers, we can see a sharp decline in the percentages of first authors (see Figure~\ref{fig:author_first_author_percentage}). Overall, early career researchers are publishing more in their careers but appear as first authors much less than in previous generations.

\subsection{Results of Journal Trends }
\label{sec:results_journal_trends}
\begin{figure*}
\centering
\subfigure{%

\includegraphics[width=0.4\linewidth, height=2.2in]{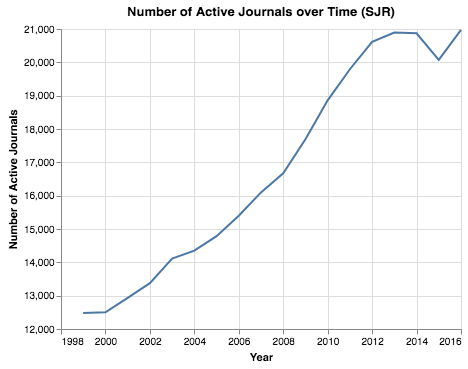}}%
\qquad
\subfigure{%
\includegraphics[ height=2.2in]{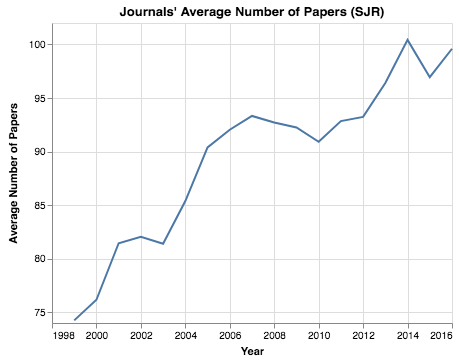}}%
\caption{\textbf{Number of Active Journals over Time.} Over a period of 18 years, from 1999 to 2016, both the number of active journals and the papers per journal increased greatly.}
\label{fig:journal_number}
\end{figure*}

By analyzing journal trends using the SJR and MAG datasets, we discovered that the number of journals increased significantly over the years, with 20,975 active ranked journals in 2016 (see Figure~\ref{fig:journal_number}). Furthermore, we observed that hundreds of new ranked journals were published each year (see Figures~\ref{fig:jornual_number_mag} and~\ref{fig:jornual_sjr_new_papers}). In addition, we discovered that the number of published papers per journal increased sharply, from an average of 74.2 papers in 1999 to an average of 99.6 papers in 2016 (see Figure~\ref{fig:journal_number}). We also observed that in recent years, journals that publish thousands of papers have become more common. For example, in 2016, according to the SJR dataset, 197 journals published over 1000 papers each.

\begin{figure}
\centering 
\includegraphics[width=\linewidth]{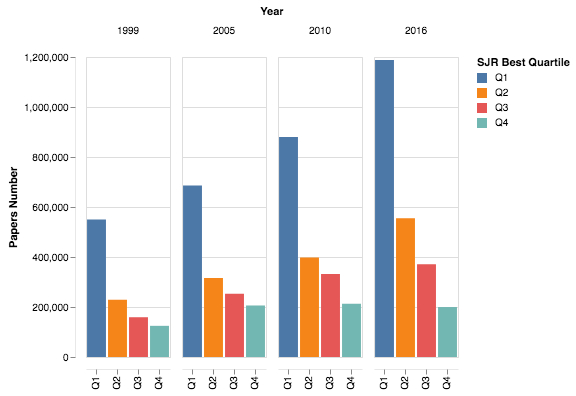}
\caption{\textbf{Journals' Quartile Number of Papers over Time.} The number of papers published in Q1 journals has vastly increased.}
\label{fig:journal_quartile}
\end{figure}

\begin{figure*}
\centering
\subfigure{%

\includegraphics[width=0.4\linewidth, height=2.2in]{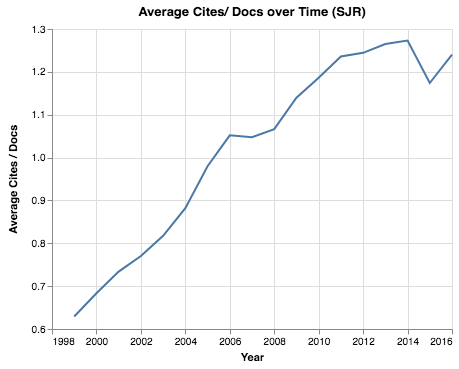}}%
\qquad
\subfigure{%
\includegraphics[ height=2.2in]{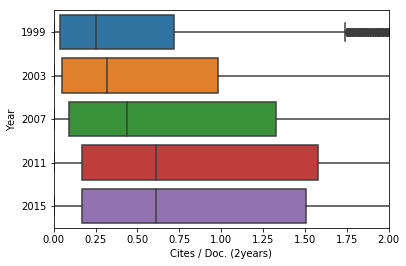}}%
\caption{\textbf{The Average Number of Citations ($\frac{Cites}{Docs}  \mbox{ (2 years)}$) over Time.} The  average number of citations values have almost doubled in the last 18 years; additionally, their distributions have changed considerably.}
\label{fig:journal_avg_cite}
\end{figure*}

By exploring how various metrics have changed over time, we discovered the following: First, over the last 18 years, the number of papers published in Q1 and Q2 journals more than doubled, from 550,109 Q1 papers and 229,373 Q2 papers in 1999, to 1,187,514 Q1 papers and 554,782 Q2 papers in 2016 (see Figure~\ref{fig:journal_quartile}). According to the SJR dataset, in 2016, 51.3\% of journal papers were published in Q1 journals and only 8.66\% were published in Q4 journals. Second, the average h-index decreased over recent years from an average value of 37.36 and median value of 23 in 1999 to an average value of 31.3 and median value of 16 in 2016 (see Figure~\ref{fig:journal_h_index}). Third, we noted that the SJR and the average number of citations measures both increased considerably during the last 18 years (see Figures~\ref{fig:journal_avg_cite} and~\ref{fig:journal_sjr_values}).

We selected the top 40 journals with the highest SJR values in 2016. We matched the journals' titles and ISSNs with the data in our Join dataset to identify the journal IDs in the MAG dataset. Using this method, we identified 30 unique journal IDs in the MAG dataset that published 110,825 papers with over 5 references. Next, we performed an in-depth analysis on these top-selected journal papers to better understand how various properties changed over time. We discovered that between 2000 and 2014, the number of papers in these journals more than doubled, and the number of authors increased significantly (see Figure~\ref{fig:journal_top_paper_author_number}).\footnote{The total number of authors each year was determined by summing the number of authors in each published paper.}  

\begin{figure}
\centering 
\includegraphics[width=\linewidth]{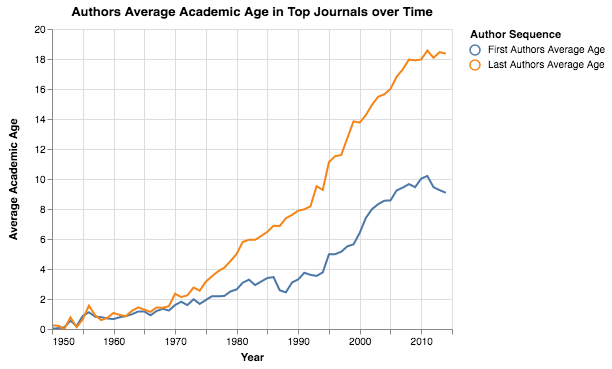}
\caption{\textbf{Top-Selected Journals' Average First and Last Authors Ages.} Both the first and last authors' average ages have increased sharply. }
\label{fig:top_journal_author_ages}
\end{figure}

\begin{figure} 
\centering 
\includegraphics[width=\linewidth]{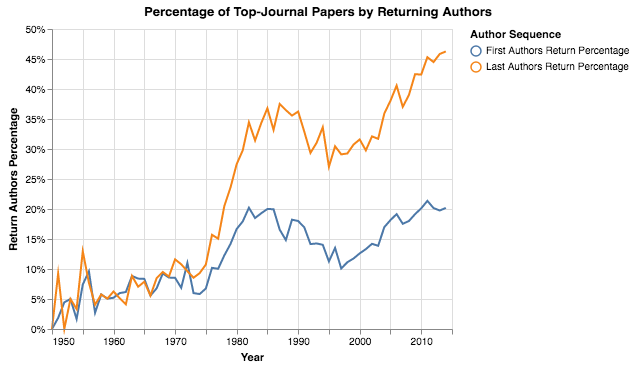}
\caption{\textbf{Percentage of Papers with Returning First or Last Authors.} The percentage of returning first or last top-journal authors increased considerably. }
\label{fig:top_journal_author_return}
\end{figure}

Additionally, by calculating the average academic career ages of first and last authors, we discovered that in recent years the average academic age has increased notably (see Figure ~\ref{fig:top_journal_author_ages}).  Moreover, when looking at first and last authors who previously published in one of the selected top-30 journals, we discovered that over time the percentage of returning authors increased substantially (see Figure~\ref{fig:top_journal_author_return}). By 2014, 46.2\% of all published papers in top-30 selected journals were published by last authors who had published at least one paper in a top-30 selected journal before (see Figure~\ref{fig:top_journal_author_return}).

\begin{figure*}[ht]
\centering 
\includegraphics[width=\linewidth]{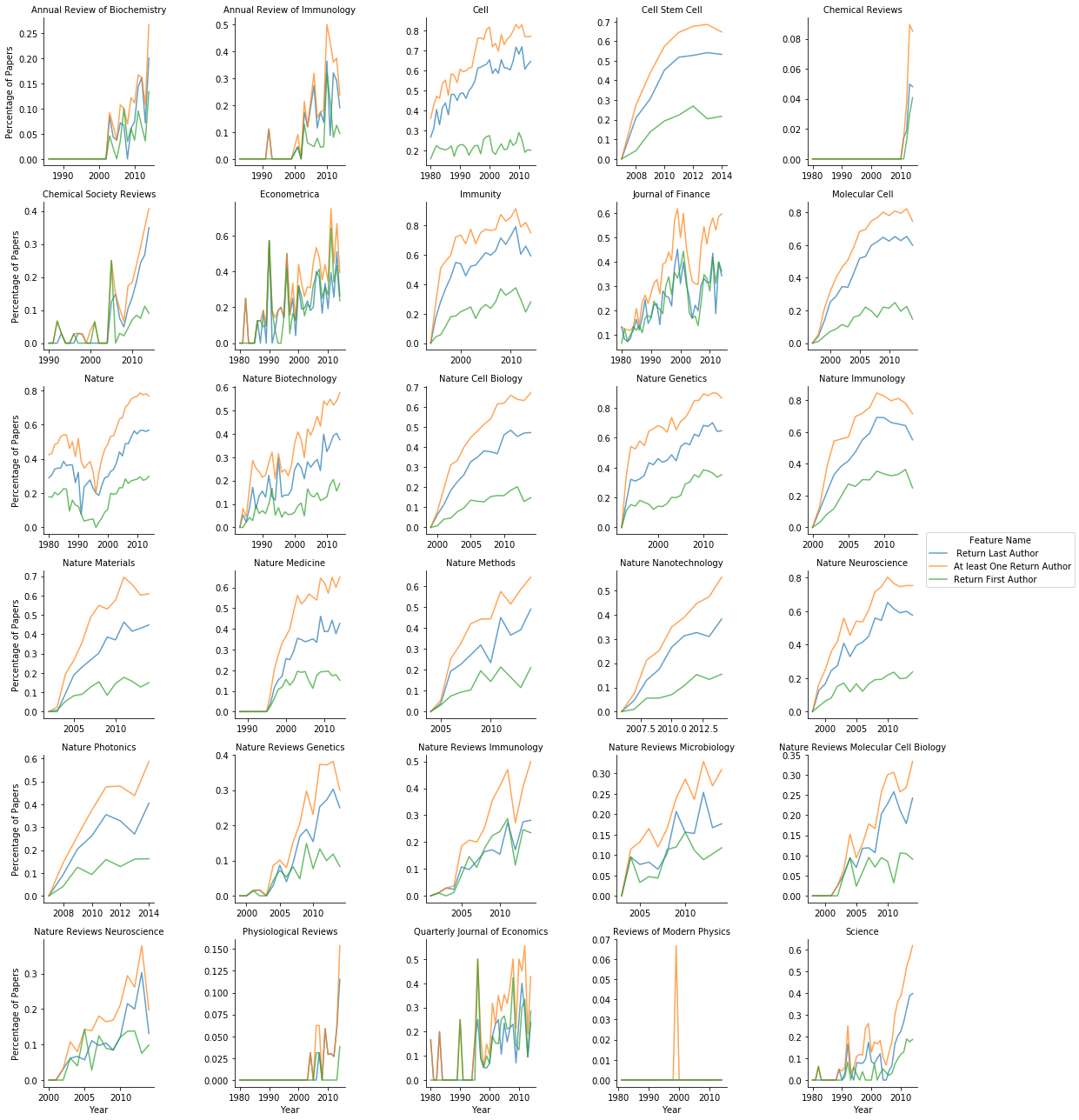}
\caption{\textbf{Average Percentage of Return Authors in Top-Selected Journals over Time.} In most journals the number of papers with at least one author who previously published in the journal increased sharply. In many of the selected journals the percentage of papers with returning authors was above 60\%, and in some cases above 80\%.} 
\label{fig:top_journal_grid_return}
\end{figure*}

By calculating the number of papers, number of authors, authors' average age, and percentage of returning authors in each selected top-30 journal, we observed the following: (a) the number of published papers per year increased considerably in the vast majority of the journals (see Figure~\ref{fig:top_journal__grid_paper_number});\footnote{Due to missing references in the MAG dataset, there are decline in the number of papers in Nature (1990s), and in Science (before 2008).} (b) the average career ages of last authors in the vast majority of the selected journals considerably increased (see Figure~\ref{fig:top_journal_grid_age}), like in Cell journal where the last authors' career ages increased from about 4.5 years in 1980 to about 20 years in 2014 (see Figure~\ref{fig:top_journal_grid_age}); and (c) the percentage of returning authors in the vast majority of the selected journals increased drastically, like in Nature Genetics where in 86.6\% of 2014 papers, at least one of the authors had published in the journal before (see Figure~\ref{fig:top_journal_grid_return}).

\subsection{Results of Fields-of-Research Trends }
\label{sec:results_fields_trends}
\begin{figure} 
\centering 
\includegraphics[width=\linewidth]{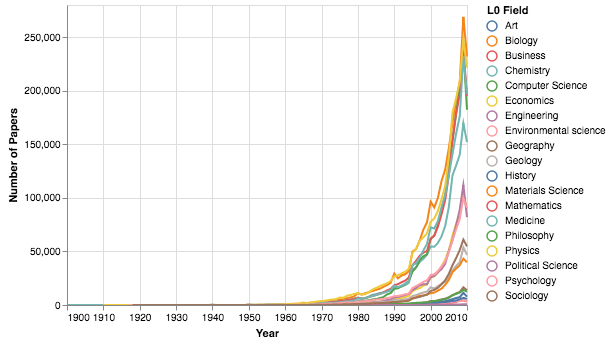}
\caption{\textbf{L0 Fields-of-Study Number of Papers over Time.} The numbers of papers in each field of study have increased drastically.}
\label{fig:field_number_papers}
\end{figure}

\begin{figure*}[ht]
\centering 
\includegraphics[width=\linewidth]{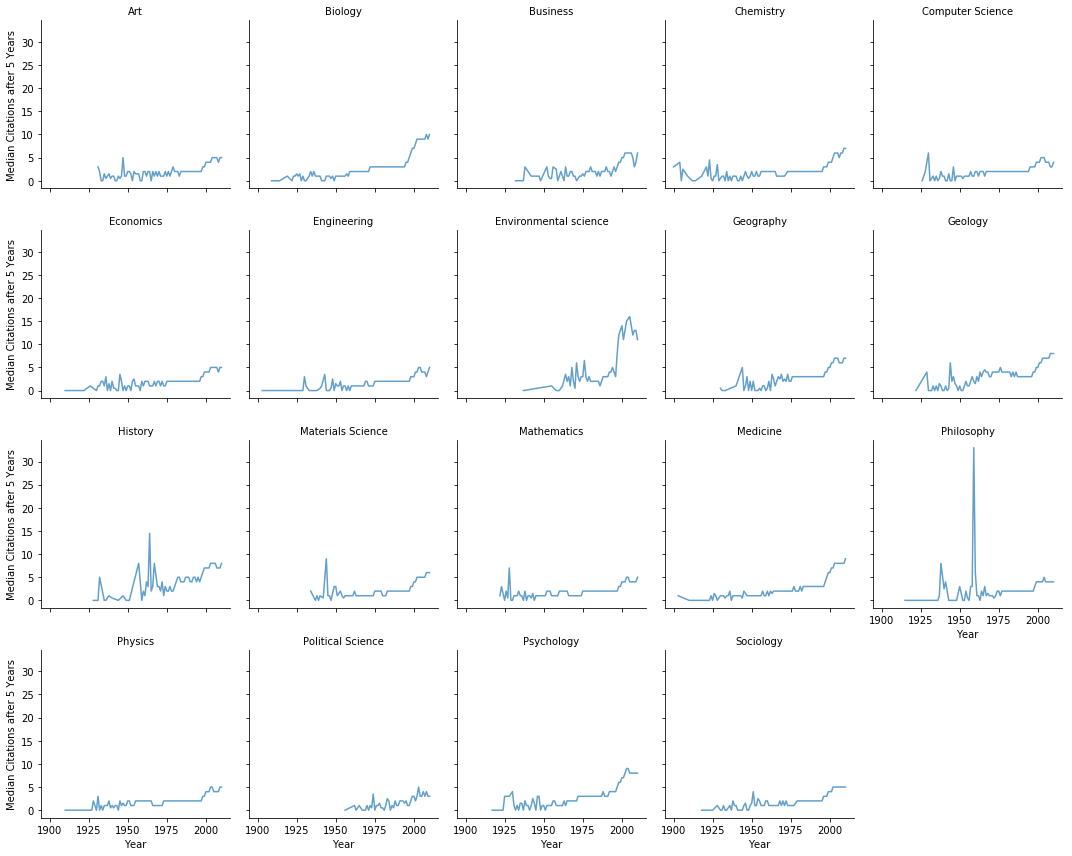}
\caption{\textbf{L0 Field-of-Study Median Citation Number after 5 Years.} There is notable variance among the L0 fields-of-study median citation numbers.}
\label{fig:field_med_citation_grid}
\end{figure*}

By matching each paper to its L0 field of study and analyzing each field's properties, we discovered substantial differences in these properties. Namely, we observed the following: 
\begin{itemize}
    \item  A large variance in the number of published papers in each field. For example, 231,756 papers were published in the field of biology in 2010, but only 5,684 were published that year in the field of history (see Figures~\ref{fig:field_number_papers} and~\ref{fig:field_num_papers_grid}).
    
    \item A considerable variance in the average number of paper authors among the various research fields. For example, the number of authors in 2010 ranged from an average of 2.28 authors in the field of political science to an average of 5.39 authors in medicine (see Figure~\ref{fig:field_num_authors_grid}).
    
    \item A variance in the papers' average number of references in different fields. For example,  in 2010, the average reference number in the fields of material science and engineering was less than 24, while in the fields of biology and history it was over 33 (see Figure~\ref{fig:field_num_ref_grid}).
    
    \item A big variance in each L0 field's average and median number of citations after 5 years. For example, for 2009 papers in the fields of computer science and political science, the median citation number after 5 years was 4 citations. In biology and environmental science, the median citation number after 5 years was 9 and 13 citations, respectively (see Figure~\ref{fig:field_med_citation_grid}).
\end{itemize}

By repeating the above analysis for the L1 subfields of biology and for the L2 subfields of genetics, we uncovered similar differences among fields of study. Namely, we observed the following for subfields in the same hierarchal level: (a) significant variance in the average number of papers (see Figures~\ref{fig:field_l1_num_papers_grid} and~\ref{fig:field_l2_num_papers_grid}); (b) notable variance in the average number of authors (see Figures~\ref{fig:field_l1_num_authors_grid} and~\ref{fig:field_l2_num_authors_grid}); (c) noteworthy variance in the average number of references (see Figures~\ref{fig:field_l1_num_ref_grid} and~\ref{fig:field_l2_num_ref_grid}); and (d) vast variance in median citation numbers (see Figures~\ref{fig:field_l1_num_citation_grid} and~\ref{fig:field_l2_num_citation_grid}).

Lastly, by analyzing various features of 2,673 L3 fields of study, we observed a huge variance in the different properties (see Table~\ref{tab:l3} and Figure~\ref{fig:fields_l3_subs}). For example, several fields of study, such as gallium (chemistry), ontology (computer science), and presentation of a group (mathematics), had median citation numbers of 2, while other fields of study, such as microRNA and genetic recombination (biology), had median citation numbers of over 47 and 50.5, respectively (see Table~\ref{tab:l3} and Figure~\ref{fig:fields_l3_subs}). 
\begin{table*}
    \centering
    \caption{L3 Fields-of-Study Features in 2009}    
\includegraphics[width=\linewidth]{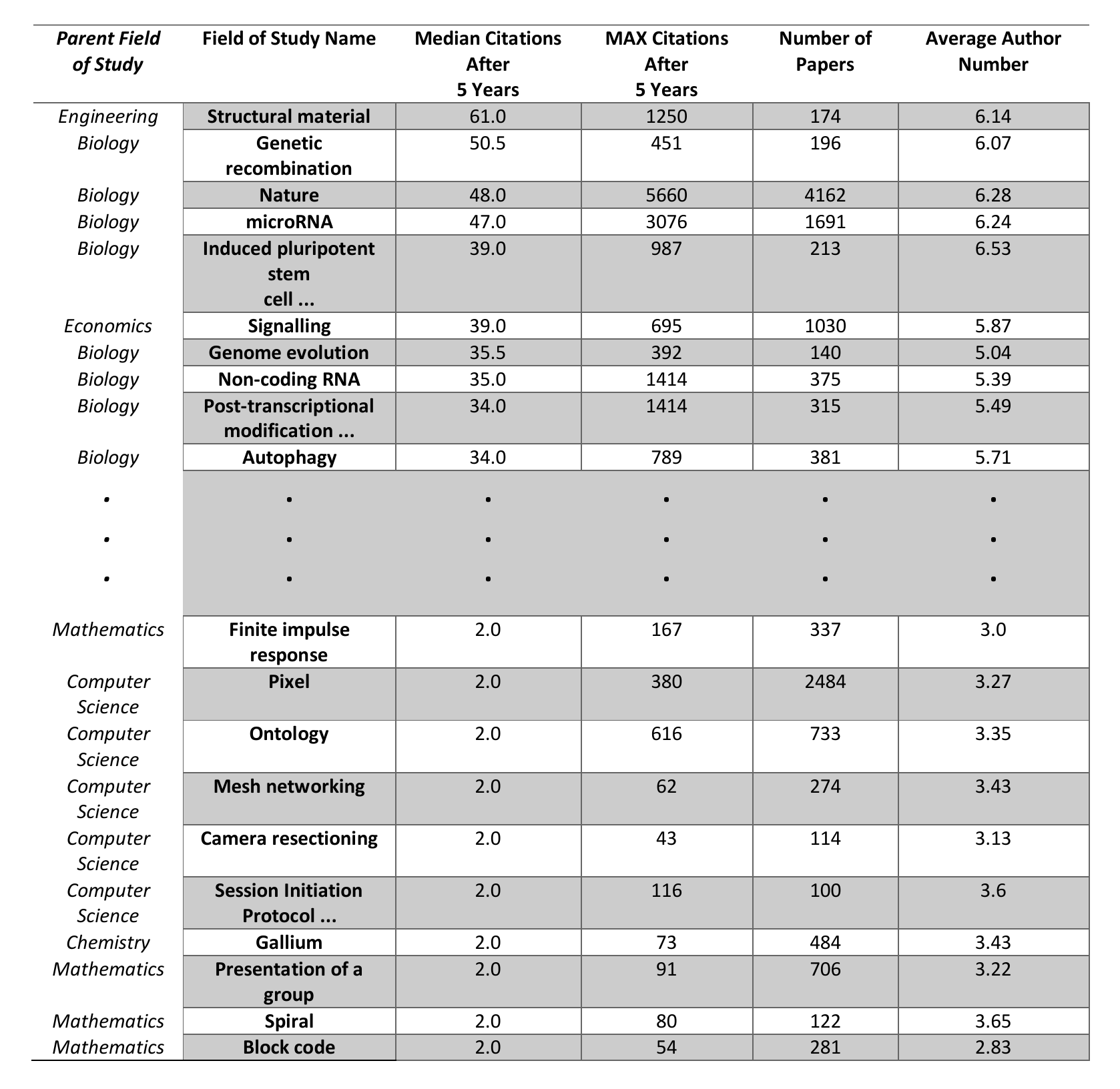}
    \label{tab:l3}
\end{table*}

\section{Discussion }
\label{sec:diss}
By analyzing the results presented in Section~\ref{sec:results}, the following can be noted: 

First, we can observe that the structure of academic papers has changed in distinct ways in recent decades. While the average overall length of papers has become shorter (see Figure~\ref{fig:paper_length}), the title, abstract, and references have become longer (see Section~\ref{sec:paper_results} and Figures~\ref{fig:title_longer},~\ref{fig:paper_abstract_len_dists},~\ref{fig:paper_authors_number},~\ref{fig:paper_avg_abstract_len},~\ref{fig:paper_avg_ref}, and~\ref{fig:paper_ref_dists}). Also, the number of papers that include keywords has increased considerably, as has the average number of keywords in each paper (see Figure~\ref{fig:paper_keywords}). Furthermore, the average and median number of authors per paper has increased sharply (see Figures~\ref{fig:paper_authors_number} and~\ref{fig:paper_max_authors}). 

These results support Goodhart's Law as it relates to academic publishing: the measures (e.g., number of papers, number of citations, h-index, and impact factor) have become targets, and now they are no longer good measures. By making papers shorter and collaborating with more authors, researchers are able to produce more papers in the same amount of time. Moreover, we can observe that the majority of changes in papers' properties are correlated with papers that receive higher numbers of citations (see Figure~\ref{fig:paper_corr}). Authors can use longer titles and abstracts, or use question or exclamation marks in titles, to make their papers more appealing. Thus more readers are attracted to the paper, and ideally they will cite it, i.e., academic clickbait~\cite{lockwood2016academic}. These results support our hypothesis that the citation number has become a target. Consequently, the properties of academic papers have evolved in order to win-to score a bullseye on the academic target. 

Second, we observed that over time fewer papers list authors alphabetically, especially papers with a relatively high number of authors (see Section~\ref{sec:paper_results} and Figures~\ref{fig:paper_authors_alpha_order_by_authors} and~\ref{fig:paper_authors_alpha_order}). These results may indicate the increased importance of an author's sequence number in a paper, which may reflect the author's contribution to the study. This result is another signal of the rising importance of measures that rate an individual's research contribution.

Third, from matching papers to their L0 fields of study, we observed that the number of multidisciplinary papers has increased sharply over time (see Figure~\ref{fig:paper_multidis}). It is important to keep in mind that these results were obtained by matching keywords to their corresponding fields of study. Therefore, these results have several limitations: First, not all papers contain keywords. Second, the dataset may not extract keywords from papers in the correct manner. For example, we found some papers contained keywords in their online version but not in their offline version (see Section~\ref{sec:paper_results}). It is also possible that in some fields it is less common to use keywords. Therefore, the papers' keywords may be missing in the datasets, and the presented results may be an underestimate of the actual number of multidisciplinary studies. Nevertheless, we observed a strong trend in increasing numbers of multidisciplinary papers.

Fourth, from seeing sharp increases in both the maximal and average number of self-citations (see Section~\ref{sec:paper_results} and Figures~\ref{fig:paper_self_avg_max},~\ref{fig:no_citations},~\ref{fig:citations_3d}, and~\ref{fig:paper_self_total_percentage}), it is clear that citation numbers have become a target for some researchers who cite their own papers dozens, or even hundreds, of times. Furthermore, we can observe a general increasing trend for researchers to cite their previous work in their new studies. Moreover, from analyzing the percentage of papers without citations after 5 years, we can observe that a huge quantity of papers -- over 72\% of all papers and 25\% of all papers with at least 5 references -- have no citations at all (see Figure~\ref{fig:no_citations}). Obviously, many resources are spent on papers with limited impact. The lack of citations may indicate that researchers are publishing more papers of poorer quality to boost their total number of papers. Additionally, by exploring papers' citation distributions (see Figure~\ref{fig:citations_3d}), we can observe that different decades have very different citation distributions. This result indicates that comparing citation records of researchers who published papers in different time periods can be challenging.

Fifth, by exploring trends in authors (see Section~\ref{sec:results_author_trends} and Figures~\ref{fig:author_pub_since_first},~\ref{fig:author_coauthors},~\ref{fig:author_first_author_percentage}, ~\ref{fig:author_number},~\ref{fig:author_journal_conf}, and~\ref{fig:author_med_seq}), we observed an exponential growth in the number of new researchers who publish papers. We also observed that young career researchers tend to publish considerably more than researchers in previous generations, using the same time frames for comparison (see Figure~\ref{fig:author_pub_since_first}). Moreover, young career researchers tend to publish their work much more in conferences in the beginning of their careers than older researchers did in previous decades (see Figure~\ref{fig:author_journal_conf}). We also observed that young career researchers tend to collaborate considerably more in the beginning of their careers than those who are older (see Figure~\ref{fig:author_coauthors}).  Furthermore, we see that the average percentage of researchers as first authors early in their career is considerably less than those in previous generations (see Figure~\ref{fig:author_first_author_percentage}). In addition, authors' median sequence numbers typically increase over time, and the rate is typically faster for young career researchers (see Figure~\ref{fig:author_med_seq}). These results emphasize the changes in academia in recent years. In a culture of ``publish or perish,'' researchers publish more by increasing collaboration (and being added to more author lists) and by publishing more conference papers than in the past. However, as can be observed by the overall decline of researchers as first authors, young career researchers may be publishing more in their careers but contributing less to each paper. The numbers can be misleading: a researcher who has 5 ``first author'' claims but has published 20 papers may be less of a true contributor than one with 4 ``first author'' claims and 10 published papers.

Sixth, by analyzing journal trends (see Section~\ref{sec:results_journal_trends}), we see a rapid increase in the number of ranked active journals in recent years (see Figure~\ref{fig:journal_number}). Moreover, on average, journals publish more papers than in the past, and dozens of journals publish over 1,000 papers each year (see Figure~\ref{fig:journal_number} and~\ref{fig:jornual_number_mag}). With the increase in the number of active journals, we observed rapid changes in impact measures: (a) the number of papers published in the first and second quartiles (Q1 and Q2) has increased sharply, and today the vast majority of papers are published in these quartiles (see Figure~\ref{fig:journal_quartile}); (b) the journals' average and median h-index has decreased sharply (see Figure~\ref{fig:jornual_sjr_new_papers}); and (c) both the SJR and the average number of citations  has increased considerably (see Figures~\ref{fig:journal_avg_cite} and~\ref{fig:journal_sjr_values}). With these significant changes, it is clear that some measures, such as the use of quartiles and the h-index, are rapidly losing meaning and value. Moreover, with the abundance of journals, researchers can ``shop around'' for a high impact journal and submit a rejected paper from one Q1 journal to another Q1 journal, time after time, and then start the review process again. These repeated reviews for the same paper wastes time, and in the long run the burden of reviewing papers several times may affect the quality of the reviews. 

There are compelling reasons to change the current system. We need to think about making all reviews open and online. We should consider the function of published journals; for that matter, is it even necessary to have journals in a world with over 20,000 journals that publish hundreds or even thousands of papers each year? We need to seriously evaluate the measures we use to judge research work. If all these measures have been devalued to being merely targets, they are no longer effective measures. Instead, they should be adapted to meet our current needs and priorities.

Seventh, by focusing on trends in selected top journals, we can observe that these journals have changed considerably in recent years (see Figures~\ref{fig:top_journal_author_ages},~\ref{fig:top_journal_author_return},~\ref{fig:top_journal_grid_return},~\ref{fig:journal_top_paper_author_number}, and~\ref{fig:top_journal__grid_paper_number}). The number of papers in the selected journals has increased sharply, along with the career age of the authors and the percentage of returning authors. The number of submissions to top journals, like Nature, have increased greatly in recent years~\cite{nature_criteria}; however, many of these journals mainly publish papers in which at least one of the authors has previously published in the journal (see Figure~\ref{fig:top_journal_author_return} and~\ref{fig:top_journal_grid_return}). We believe that this situation is also a result of Goodhart's Law. The target is the impact factor, and so researchers are vigorously seeking journals with high impact factors. Therefore, the yearly volume of papers sent to these top journals has considerably increased, and overwhelmed by the volume of submissions, editors at these journals may choose safety over risk and select papers written by only well-known, experienced researchers. 

Eighth, by analyzing how features evolve in the various L0 fields of study using the MAG dataset, we can observe that different fields have completely different sets of features (see Figures~\ref{fig:field_number_papers},~\ref{fig:field_med_citation_grid},~\ref{fig:field_number_papers},~\ref{fig:field_num_authors_grid},~\ref{fig:field_num_ref_grid}, and Table~\ref{tab:l3}). While some fields have hundreds of thousands of papers published yearly, others have only thousands published yearly (see Figures~\ref{fig:field_number_papers}and~\ref{fig:top_journal__grid_paper_number}). Moreover, similar large differences are reflected in other examined fields’ features, such as the average number of references and the average and median citation numbers (see Figures~\ref{fig:field_med_citation_grid} and~\ref{fig:fields_l3_subs}).

Lastly, by examining over 2600 research fields of various scales (see Table~\ref{tab:l3} and Figure~\ref{fig:fields_l3_subs}), we observed vast diversity in the properties of papers in different domains -- some research domains grew phenomenally while others did not. Even research domains in the same subfields presented a wide range of properties, including papers'
number of references and median number of citations per research field (see Table 1 and Figures~\ref{fig:field_l1_num_ref_grid},~\ref{fig:field_l2_num_ref_grid},~\ref{fig:field_l1_num_citation_grid}, and~\ref{fig:field_l2_num_citation_grid}). These results indicate that using measures such as citation number, h-index, and impact factor are useless when comparing researchers in different fields, and even for comparing researchers in the same subfield, such as genetics. These results emphasize that using citation-based measures for comparing various academic entities is like comparing apples to oranges, and is to ``discriminate between scientists.''~\cite{lehmann2006measures}. Moreover, using these measures as gauges to compare academic entities can drastically affect the allocation of resources and consequently damage research. For example, to improve their world ranking, universities might choose to invest in faculty for computer science and biology, rather than faculty for less-cited research fields, such as economics and psychology. Moreover, even within a department, the selection of new faculty members can be biased due to using targeted measures, such as citation number and impact factor. A biology department might hire genetic researchers in the field of epigenetics, instead of researchers in the field of medical genetics, due to the higher average number of citations in the epigenetics field. Over time, this can unfairly favor high-citation research fields at the expense of other equally significant fields.

\section{Conclusions}
\label{sec:conclusions}
In this study, we performed a large-scale analysis of academic publishing trends, utilizing data on over 120 million papers and over 20,000 journals. By analyzing this huge dataset, we can observe that over the last century, especially the last few decades, published research has changed considerably, including the numbers of papers, authors, and journals; the lengths of papers; and the average number of references in specific fields of study. 

While the research environment has changed, the measures to determine the impact of papers, authors, and journals have not changed. Measures based on citations, such as impact factor and citation number, were used 60 years ago, in a time before preprint repositories and mega-journals existed and before academia became such a hypercompetitive environment. Most important, however, is that these measures have degenerated into becoming purely targets. Goodhart's Law is clearly being illustrated: when a citation-based measure becomes the target, the measure itself ceases to be meaningful, useful, or accurate.

Our study's extensive analysis of academic publications reveals why using citation-based metrics as measures of impact are wrong from the core: First, not all citations are equal; there is a big difference between a study that cites a paper that greatly influenced it and a study that cites multiple papers with only minor connections. Many of the impact measures used today do not take into consideration distinctions among the various types of citations. Second, it is not logical to measure a paper's impact based on the citation numbers of other papers that are published in the same journal. In the academic world, there are over 20,000 journals that publish hundreds or even thousands of papers each year, with papers written by hundreds or even thousands of authors. It is even less logical to measure a researcher's impact based on a paper coauthored with many other researchers according to the journal in which it is published. Third, as we demonstrated in Section~\ref{sec:results_fields_trends}, it is wrong to compare studies from different fields, and even to compare papers and researchers within the same parent field of study, due to the many differences in the median and average number of citations in each field (see Table~\ref{tab:l3}). 

As we have revealed in this study, to measure impact with citation-based measures—that have now become targets—clearly has many undesirable effects. The number of papers with limited impact has increased sharply (see Figure~\ref{fig:no_citation_total}), papers may contain hundreds of self-citations (see Figure~\ref{fig:paper_self_avg_max}), and some top journals have become ``old boys' clubs'' that mainly publish papers from the same researchers (see Figures~\ref{fig:top_journal_author_ages} and~\ref{fig:top_journal_author_return}). Moreover, using citation-based measures to compare researchers in different fields may have the dangerous effect of allocating more resources to high-citation domains, shortchanging other domains that are equally important.

We believe the solution to the above issues is to utilize data-science tools and release new and open datasets in order to develop new measures that will more accurately determine a paper's impact in a specific research field. Certain metrics have been proposed, but the key is to wisely and carefully evaluate new measures to ensure that they will not follow Goodhart's Law and end up merely as targets. Researchers do valuable work. Communicating the work to others is vital, and correctly assessing the impact of that work is essential. 

\section{Data and Code Availability}
\label{sec:code}
One of the main goals of this study was to create an open source framework, which provided an easy way to query the datasets. Our code framework, including tutorials, is available at the \href{http://sciencedynamics.cs.washington.edu/}{project's website}, which also gives researchers the ability to interactively explore and better understand how various journals' properties have changed over time (see Figure~\ref{fig:website}).

\phantomsection
\section*{Acknowledgments} 
First and foremost, we would like to thank the AMiner, Microsoft Academic Graph, and SJR teams for making their datasets available online. Additionally, we thank the AWS Cloud Credits for Research. We also thank the Washington Research Foundation Fund for Innovation in Data-Intensive Discovery, the Moore/Sloan Data Science Environments Project at the University of Washington, and Microsoft Azure Research Award for supporting this study. Datasets, software implementations, code tutorials, and an interactive web interface for investigating the studied networks are available at the following \href{http://sciencedynamics.cs.washington.edu/}{link}.

We also wish to especially thank Carol Teegarden for editing and proofreading this article to completion, and to Sean McNaughton for designing and illustrating the article's infographic.

\addcontentsline{toc}{section}{Acknowledgments} 

\phantomsection
\bibliographystyle{unsrt}
\bibliography{sample}

\newpage
\appendix 
\beginsupplement
\section{Supplementary Materials - Additional Results}
\subsection{Additional Results}
\begin{figure*}\centering 
\includegraphics[width=\linewidth]{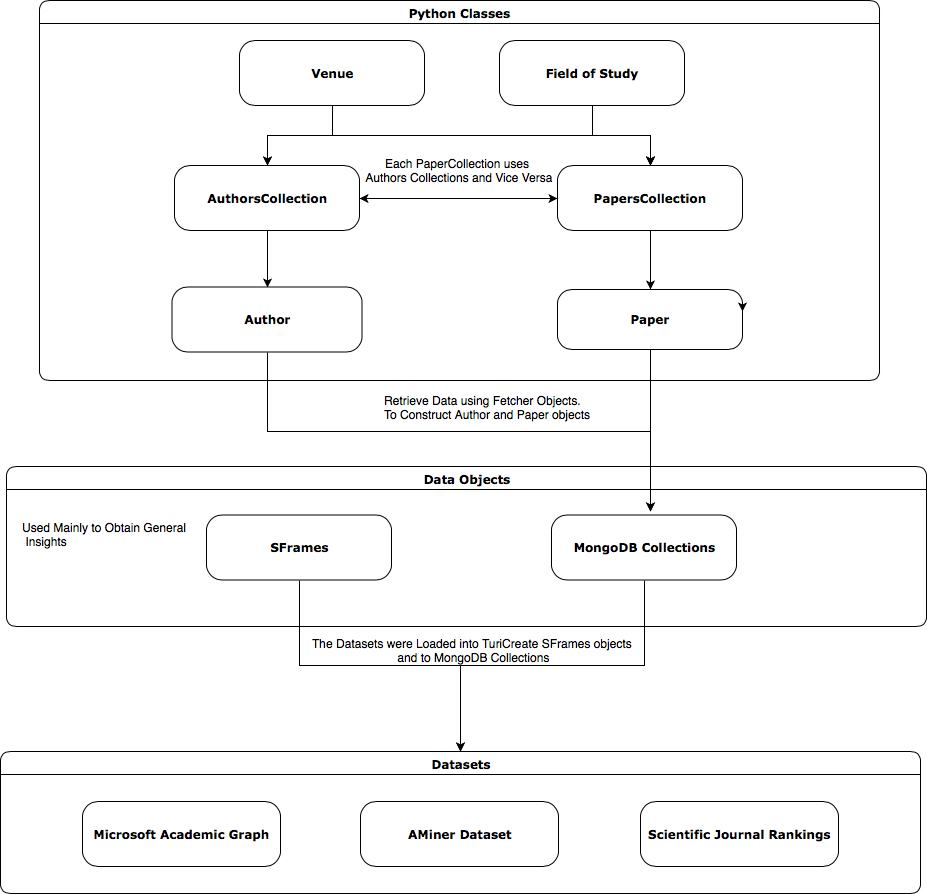}
\caption{\textbf{Overview of the Code Framework.} The datasets are loaded into SFrame objects and MongoDB collections. The SFrame objects are used mainly to obtain general insights by analyzing tens of millions of papers and author records. The MongoDB collections are used to construct Paper and Author objects that can be used to analyze more complicated statistics for specific venues and research fields with usually hundreds of thousands of records.}
\label{fig:framework}
\end{figure*}

\begin{figure*}%
\centering
\subfigure{%

\includegraphics[width=0.45\linewidth]{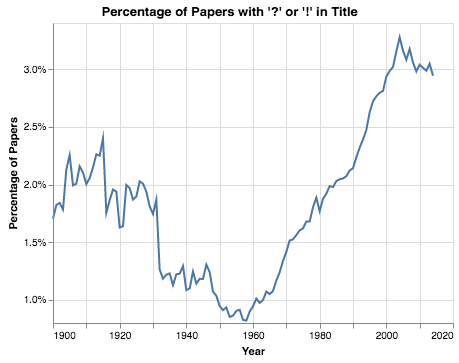}}%
\qquad
\subfigure{%
\includegraphics[width=0.45\linewidth]{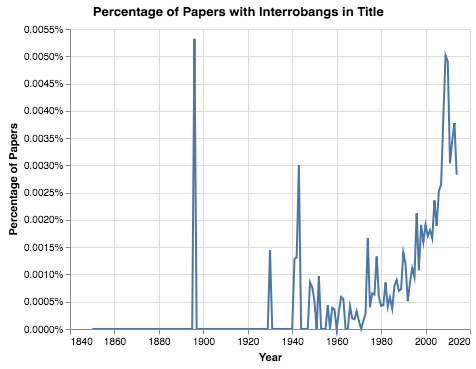}}%
\caption{\textbf{Percentage of Titles with Question or Exclamation Marks.} The percentage of papers with question or exclamation marks in their titles increased over time, as well as the percentage of titles with interrobangs (represented by ?! or !?). }
\label{fig:title_marks}
\end{figure*}

\begin{figure}
\centering 
\includegraphics[width=\linewidth]{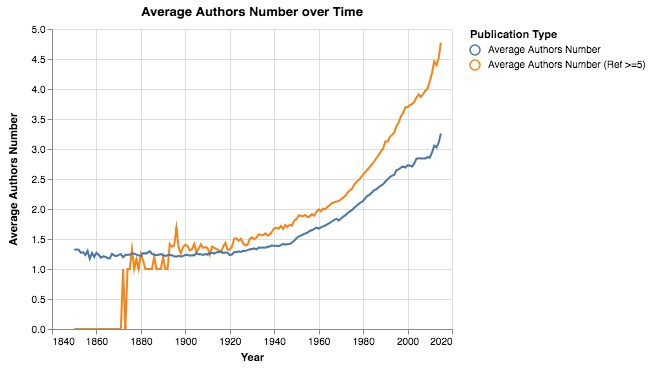}
\caption{\textbf{Average Number of Authors over Time.} There has been a rise in the average number of authors, especially in recent decades.}
\label{fig:paper_authors_number}
\end{figure}

\begin{figure}
\centering 
\includegraphics[width=\linewidth]{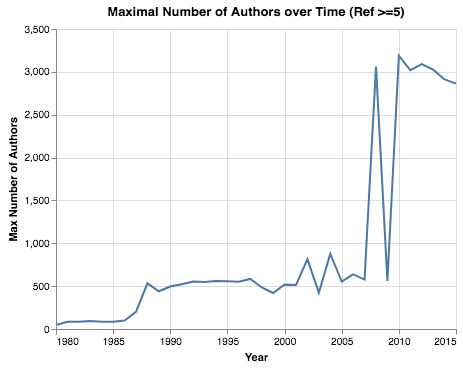}
\caption{\textbf{Maximal Number of Authors over Time.} In recent years the maximal number of authors per paper increased sharply from 520 authors in 2000 to over 3100 authors in 2010.}
\label{fig:paper_max_authors}
\end{figure}

\begin{figure}[ht]
\centering 
\includegraphics[width=\linewidth]{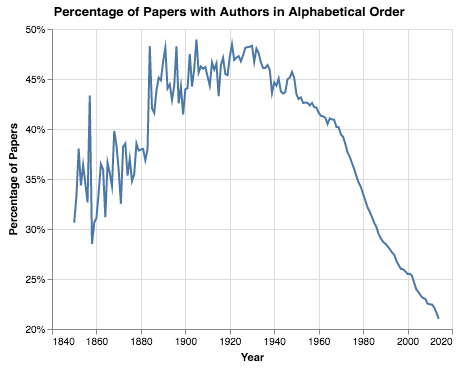}
\caption{\textbf{Percentage of Author Lists in Alphabetical Order.} There has been a decline in the number of author lists organized in alphabetical order.}
\label{fig:paper_authors_alpha_order}
\end{figure}

\begin{figure}[ht]
\centering 
\includegraphics[width=\linewidth]{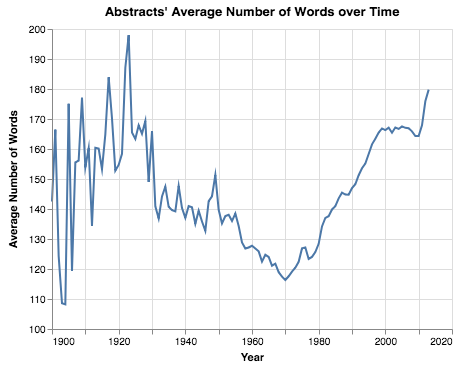}
\caption{\textbf{Average Length of Abstracts.} Since 1970 there has been an increase in abstracts' average number of words.}
\label{fig:paper_avg_abstract_len}
\end{figure}

\begin{figure*}[ht]
\centering
\subfigure{%

\includegraphics[width=0.45\linewidth]{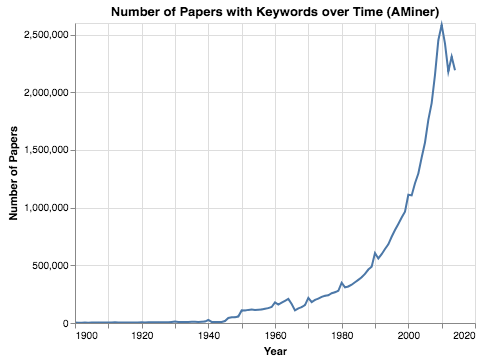}}%
\qquad
\subfigure{%
\includegraphics[width=0.45\linewidth]{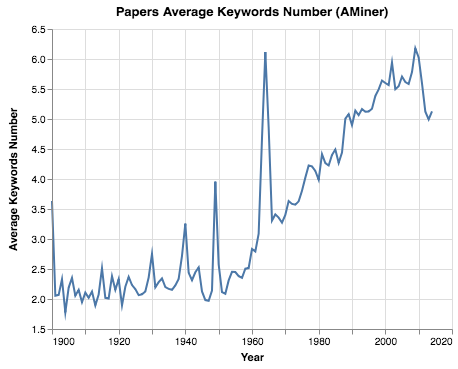}}%
\caption{\textbf{Keyword Trends.} Both the number of papers with keywords has increased, as well as the average number of keywords per paper. }
\label{fig:paper_keywords}
\end{figure*}

\begin{figure*}[ht]%
\centering
\subfigure{%

\includegraphics[width=0.45\linewidth]{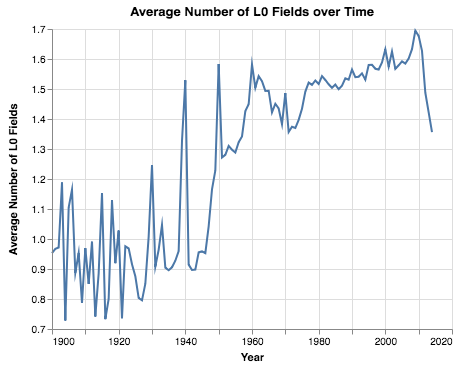}}%
\qquad
\subfigure{%
\includegraphics[width=0.45\linewidth]{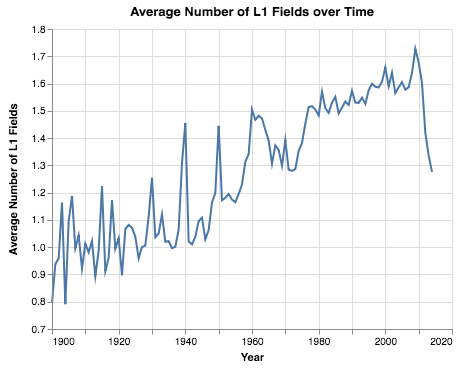}}%
\caption{\textbf{Average Number of Fields of Study over Time.} Over time both the average number of L0 and L1 fields of studies per paper considerably increased. We believe the drop in the average number of L0 and L1 fields is a direct results of the drop in the number of papers with keywords in the same years (see Section~\ref{sec:paper_results}). }
\label{fig:paper_multidis_l0_l1}
\end{figure*}

\begin{figure}[ht]
\centering 
\includegraphics[width=\linewidth]{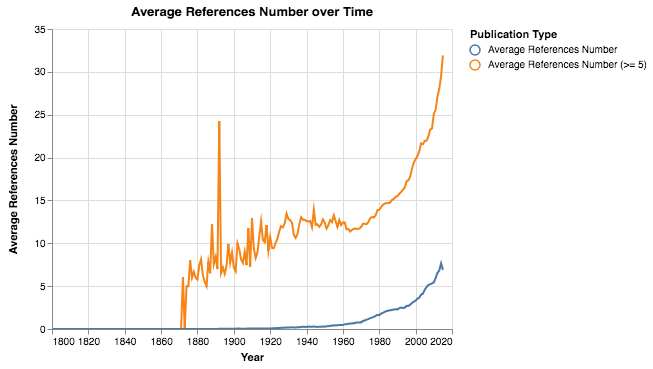}
\caption{\textbf{Average Number of References over Time.} Over time, the average number of references sharply increased.}
\label{fig:paper_avg_ref}
\end{figure}

\begin{figure*}[ht]
\centering 
\includegraphics[width=\linewidth]{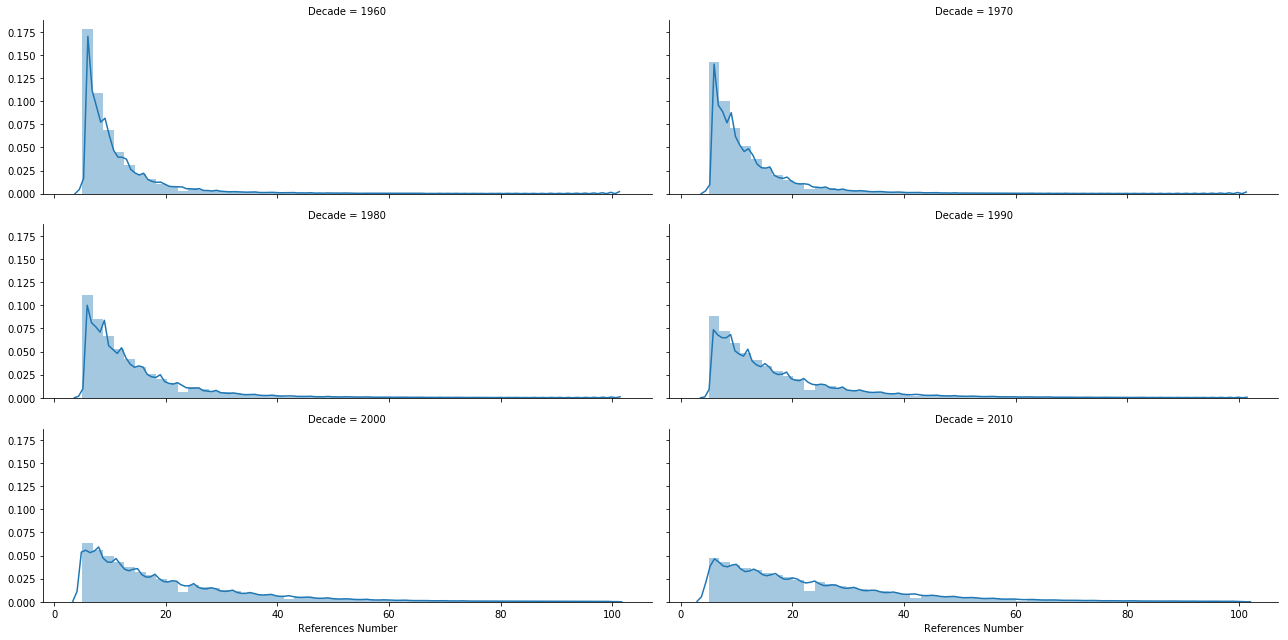}
\caption{\textbf{Distributions over Time of References in Papers.} Over time, papers with a relatively high number of references have become more common.}
\label{fig:paper_ref_dists}
\end{figure*}

\begin{figure*}%
\centering
\subfigure{%

\includegraphics[width=0.45\linewidth]{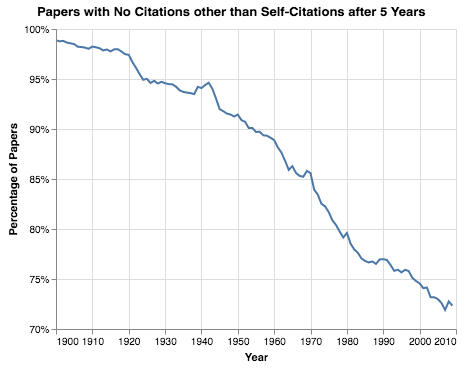}}%
\qquad
\subfigure{%
\includegraphics[width=0.45\linewidth]{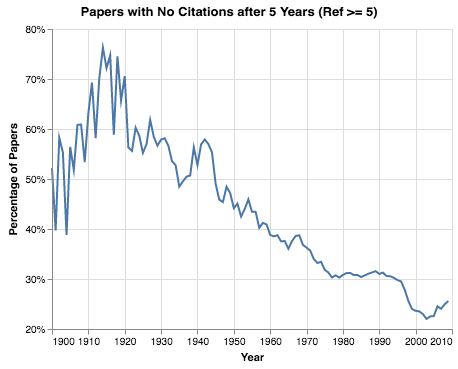}}%
\caption{\textbf{Papers with No Citations after 5 Years.} Papers with no citations after 5-years decreased; nevertheless, in 2009 over 72.1\% of all published papers had no citations after 5 years. }
\label{fig:no_citations}
\end{figure*}

\begin{figure*}[ht]%
\centering
\subfigure{%

\includegraphics[width=0.45\linewidth]{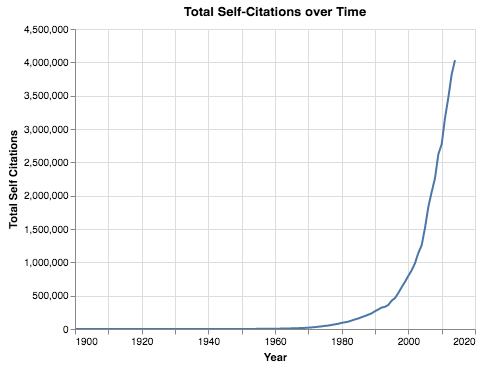}}%
\qquad
\subfigure{%
\includegraphics[width=0.45\linewidth]{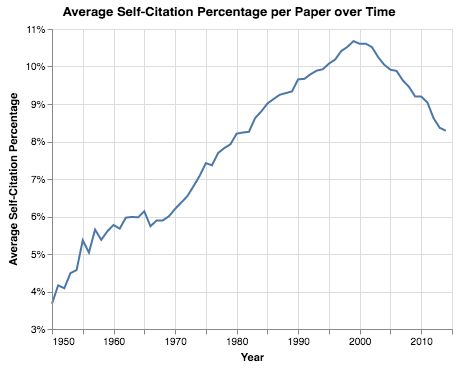}}%
\caption{\textbf{Total Number of Self-Citations and Percentage of Papers with Self-Citations.} We can observe that over time both the total number of self-citations as well as the percentage of papers with self-citations increased significantly.}
\label{fig:paper_self_total_percentage}
\end{figure*}

\begin{figure*}[ht]
\centering 
\includegraphics[width=0.8\linewidth]{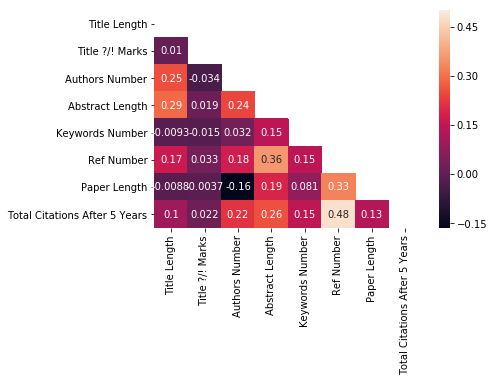}
\caption{\textbf{Spearman Correlation Heat Map for Papers' Properties.} We can observe positive correlations among papers' various structural properties and the papers' total number of citations after 5 years. }
\label{fig:paper_corr}
\end{figure*}

\begin{figure}[ht]
\centering 
\includegraphics[width=0.8\linewidth]{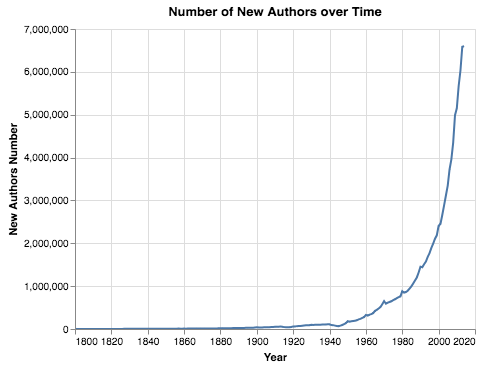}
\caption{\textbf{New Authors over Time.} The number of authors, with unique MAG author IDs, who published their first paper each year.}
\label{fig:author_number}
\end{figure}

\begin{figure*}[ht]%
\centering
\subfigure{%

\includegraphics[width=0.4\linewidth, height=2.2in]{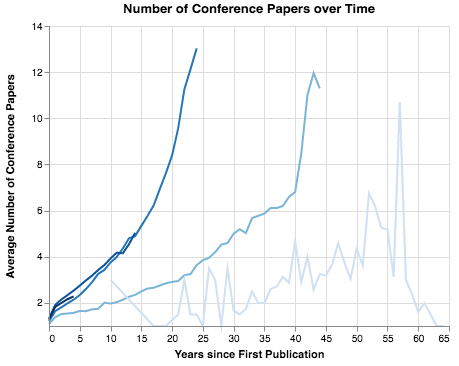}}%
\qquad
\subfigure{%
\includegraphics[ height=2.2in]{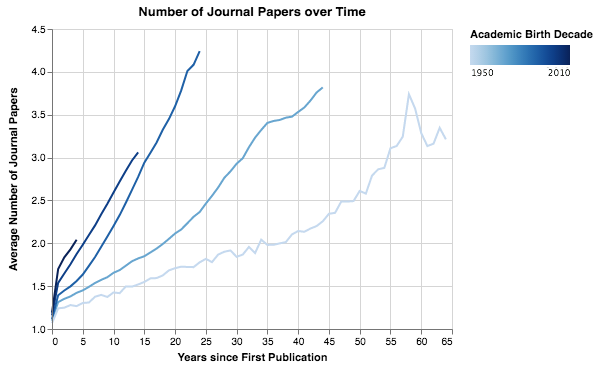}}%
\caption{\textbf{Authors Average Number of Conference and Journal Papers over Time.} The average publication rate of both journal and conference papers increased with every decade.}
\label{fig:author_journal_conf}
\end{figure*}

\begin{figure}[ht]
\centering 
\includegraphics[width=\linewidth]{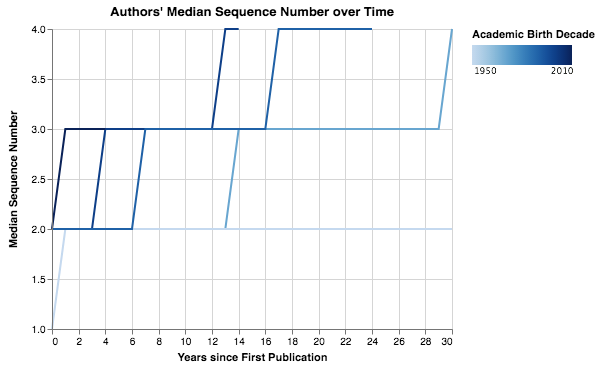}
\caption{\textbf{Authors' Median Sequence Number over Time.} We can see that over time the median sequence numbers increased; i.e., senior researchers tend to have higher sequence numbers. }
\label{fig:author_med_seq}
\end{figure}

\begin{figure}[ht]
\centering 
\includegraphics[width=\linewidth]{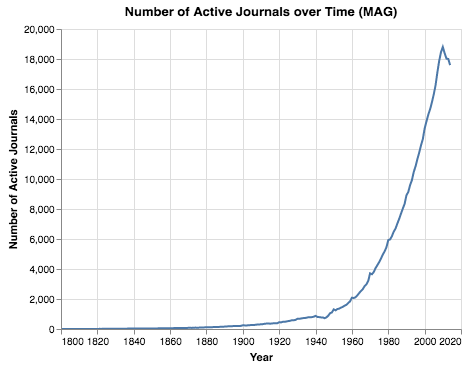}
\caption{\textbf{Number of Journals over Time according to the MAG Dataset.} There has been a drastic increase in the number of journals since the 1960s. }
\label{fig:jornual_number_mag}
\end{figure}

\begin{figure}[ht]
\centering 
\includegraphics[width=\linewidth]{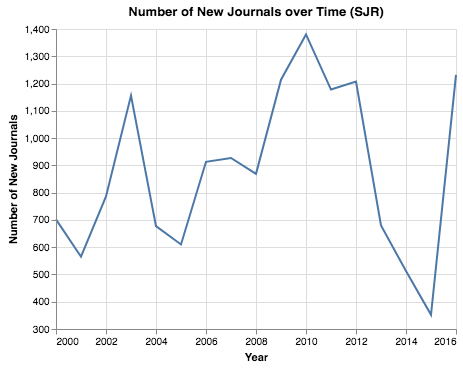}
\caption{\textbf{Number of New Journals by Year.} Hundreds of new ranked journals are being published each year. }
\label{fig:jornual_sjr_new_papers}
\end{figure}

\begin{figure*}[ht]%
\centering
\subfigure{%

\includegraphics[width=0.4\linewidth, height=2.2in]{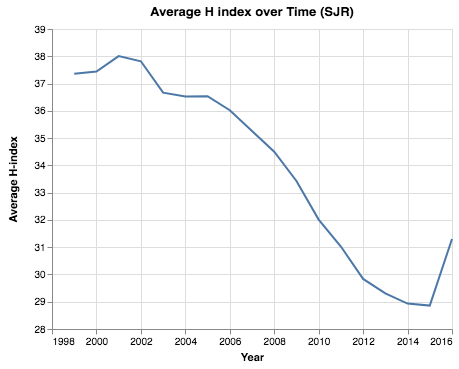}}%
\qquad
\subfigure{%
\includegraphics[ height=2.2in]{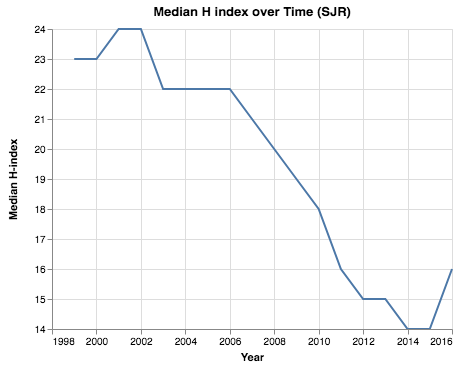}}%
\caption{\textbf{Journals' H-Index Average and Median Values.} We can notice that over time both the average and median values of the journals' h-index measures decreased.}
\label{fig:journal_h_index}
\end{figure*}

\begin{figure*}[ht]%
\centering
\subfigure{%

\includegraphics[width=0.4\linewidth, height=2.2in]{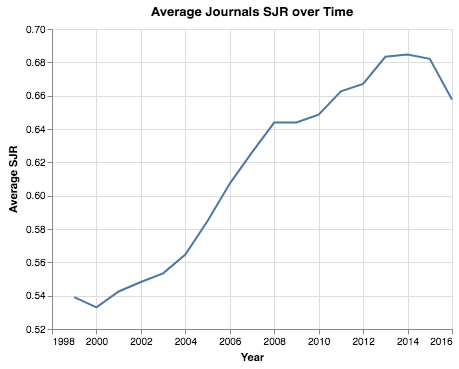}}%
\qquad
\subfigure{%
\includegraphics[ height=2.2in]{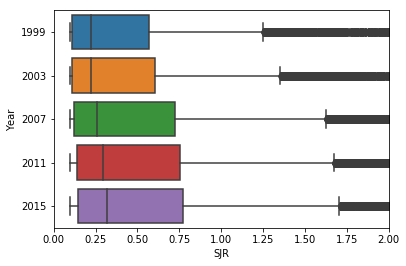}}%
\caption{\textbf{SJR Values over Time.}  We can observe that over time both the average and median SJR values increased.}
\label{fig:journal_sjr_values}
\end{figure*}

\begin{figure*}[ht]%
\centering
\subfigure{%

\includegraphics[width=0.4\linewidth, height=2.2in]{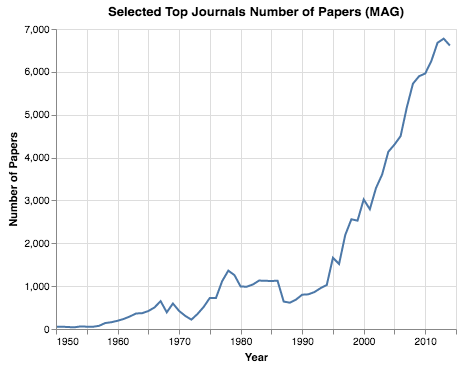}}%
\qquad
\subfigure{%
\includegraphics[ height=2.2in]{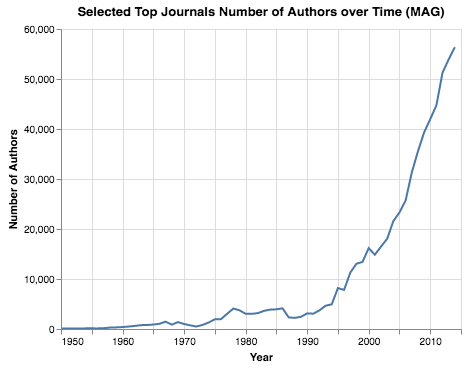}}%
\caption{\textbf{Top Journals' Number of Papers and Authors over Time.} We can observe that both the number of papers and authors increased sharply in recent years.}
\label{fig:journal_top_paper_author_number}
\end{figure*}

\begin{figure*}[ht]
\centering 
\includegraphics[width=\linewidth]{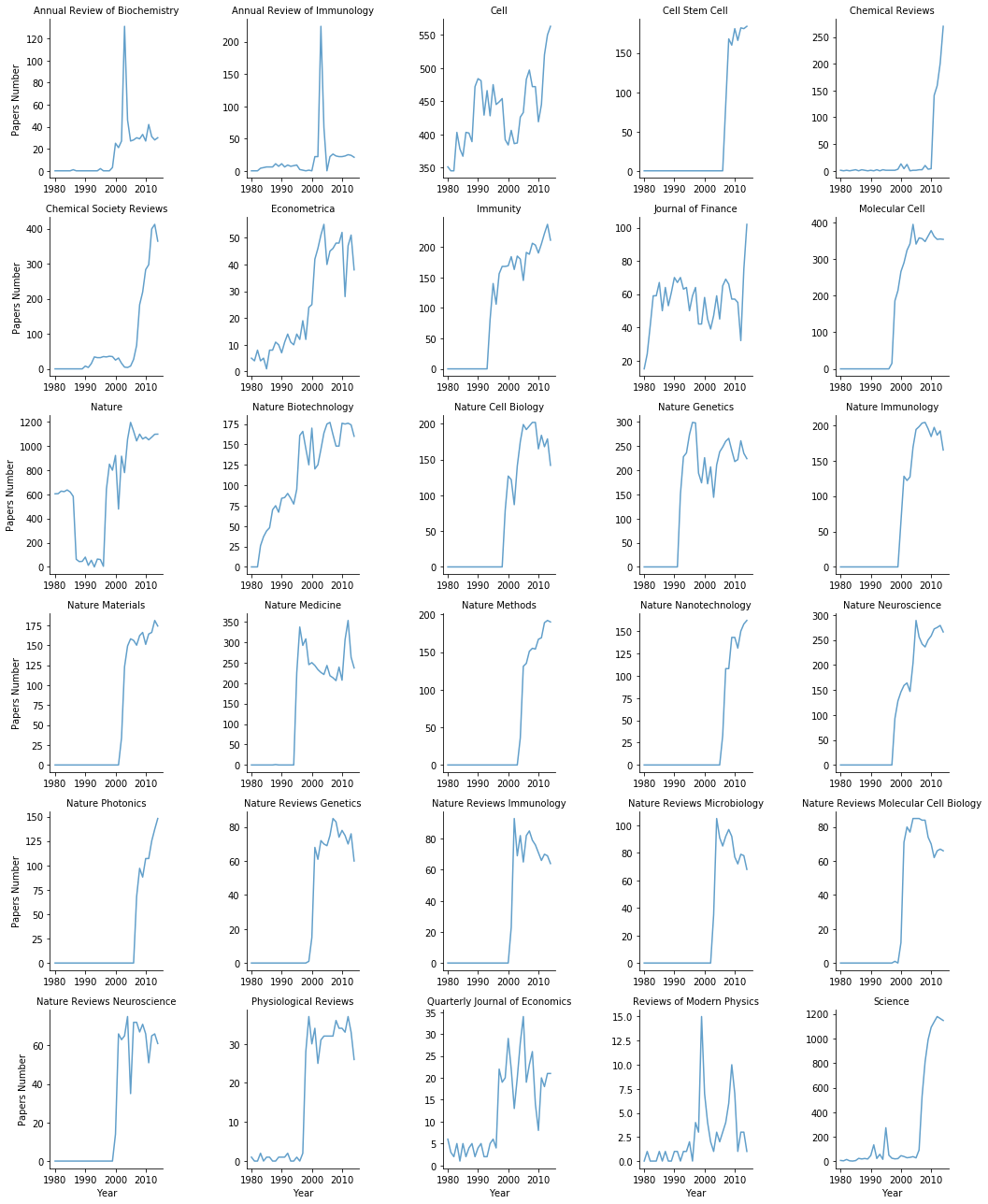}
\caption{\textbf{Top Selected Journals' Number of Papers over Time.} It can be noted that in the vast majority of the selected journals the number of published papers with at least 5 references increased considerably over time. }
\label{fig:top_journal__grid_paper_number}
\end{figure*}

\begin{figure*}[ht]
\centering 
\includegraphics[width=\linewidth]{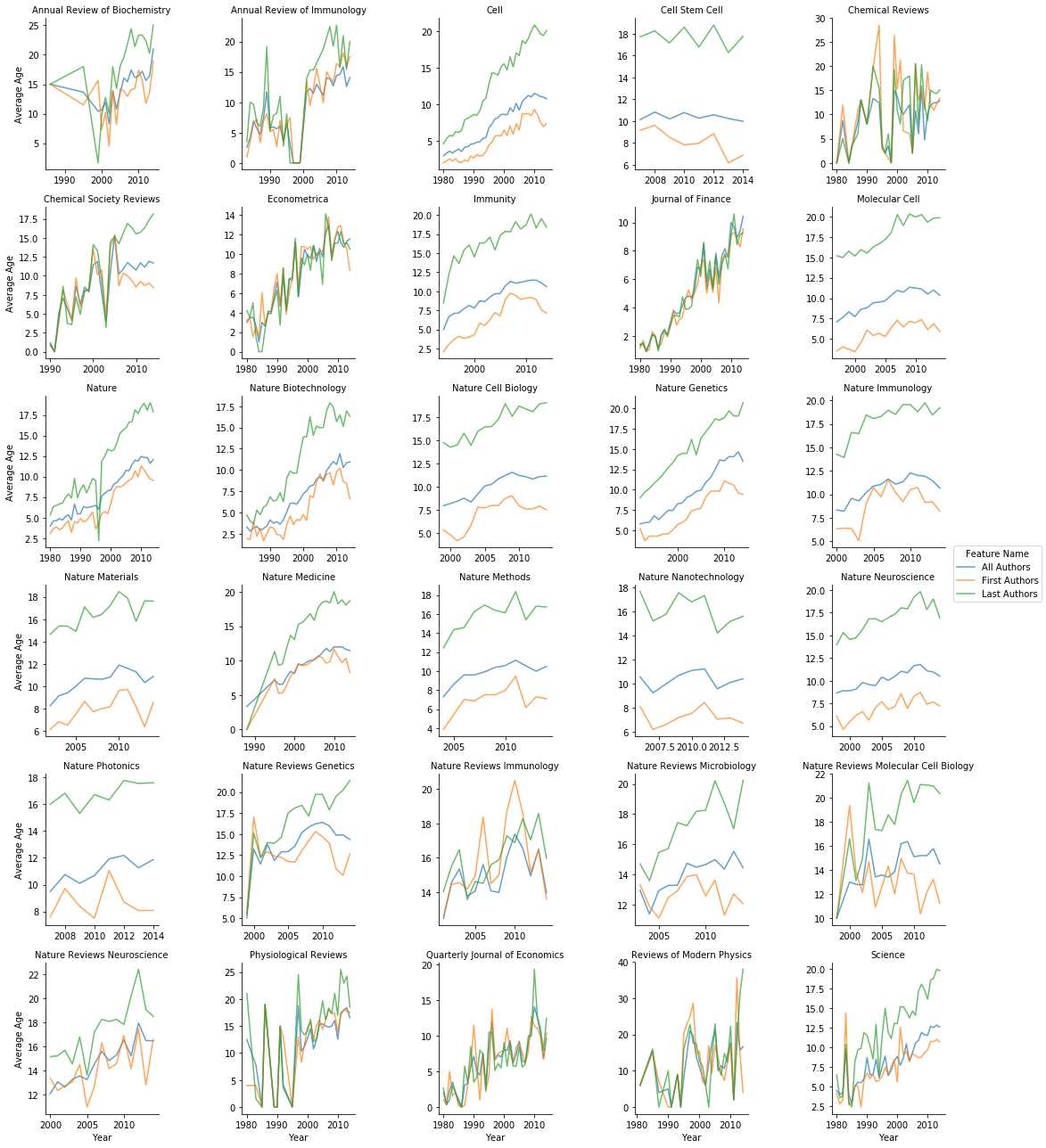}
\caption{\textbf{Top Selected Journals Average Author Career Age over Time.} It can be noted that in the vast majority of the selected journals, the average age of authors, especially last authors, increased greatly over time. }
\label{fig:top_journal_grid_age}
\end{figure*}

\begin{figure*}[ht]
\centering 
\includegraphics[width=\linewidth]{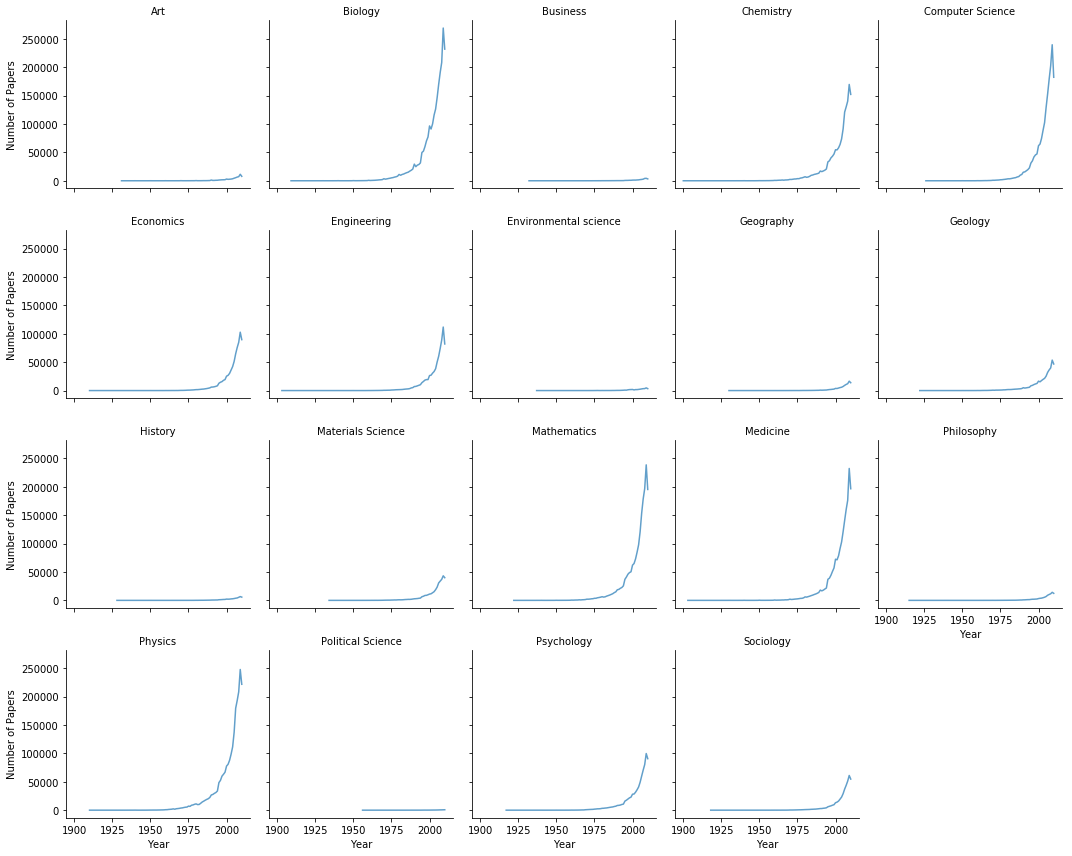}
\caption{\textbf{L0 Fields-of-Study Number of Papers over Time.} We can observe the large diversity in the number of papers published in each L0 research field.  }
\label{fig:field_num_papers_grid}
\end{figure*}

\begin{figure*}[ht]
\centering 
\includegraphics[width=\linewidth]{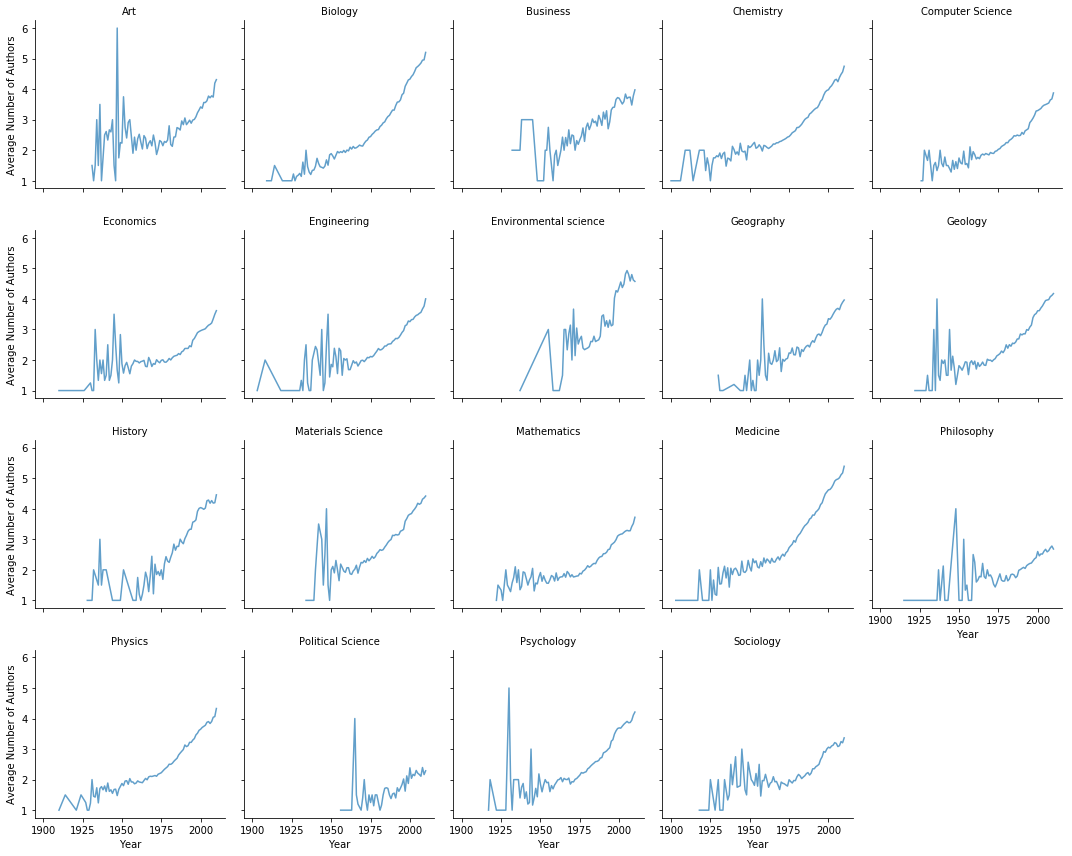}
\caption{\textbf{L0 Fields-of-Study Average Authors Number.} We can observe a variation in the average number of authors across the various research fields. }
\label{fig:field_num_authors_grid}
\end{figure*}

\begin{figure*}[ht]
\centering 
\includegraphics[width=\linewidth]{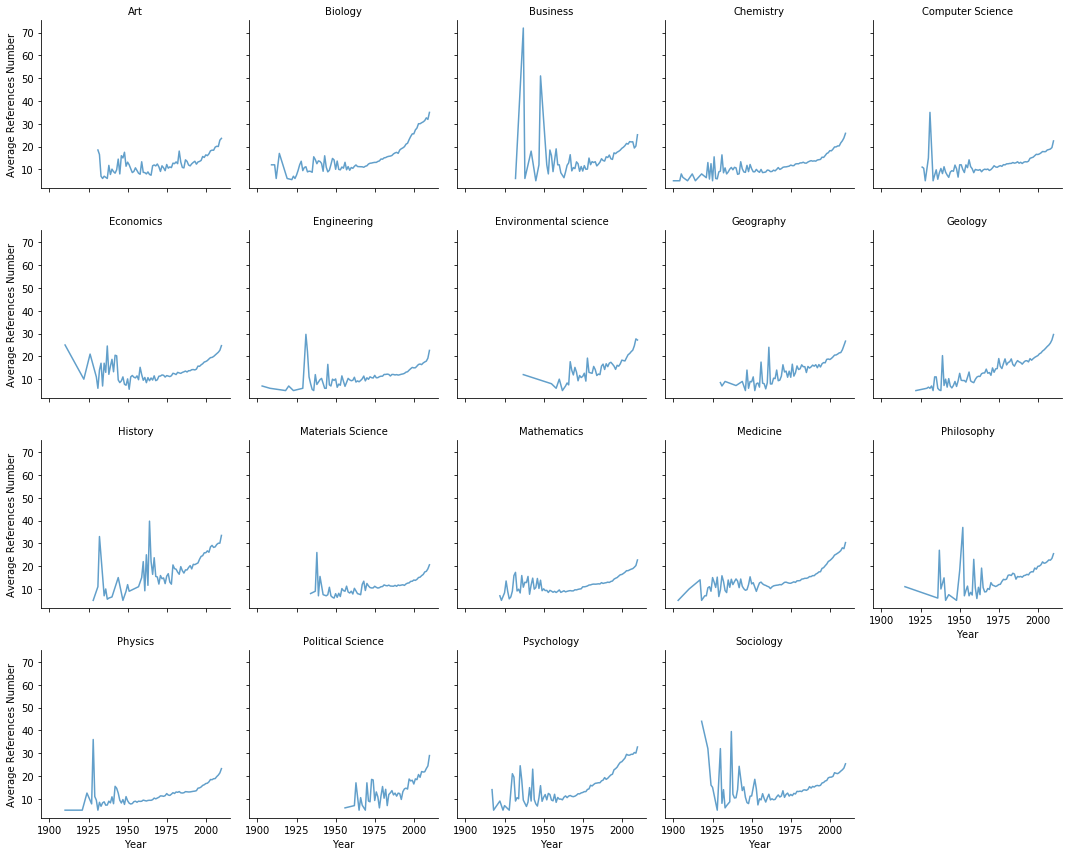}
\caption{\textbf{L0 Fields-of-Study Average References Numbers.}  We can observe variance among the reference numbers in different fields.  }
\label{fig:field_num_ref_grid}
\end{figure*}

\begin{figure*}[ht]
\centering 
\includegraphics[width=0.8\linewidth]{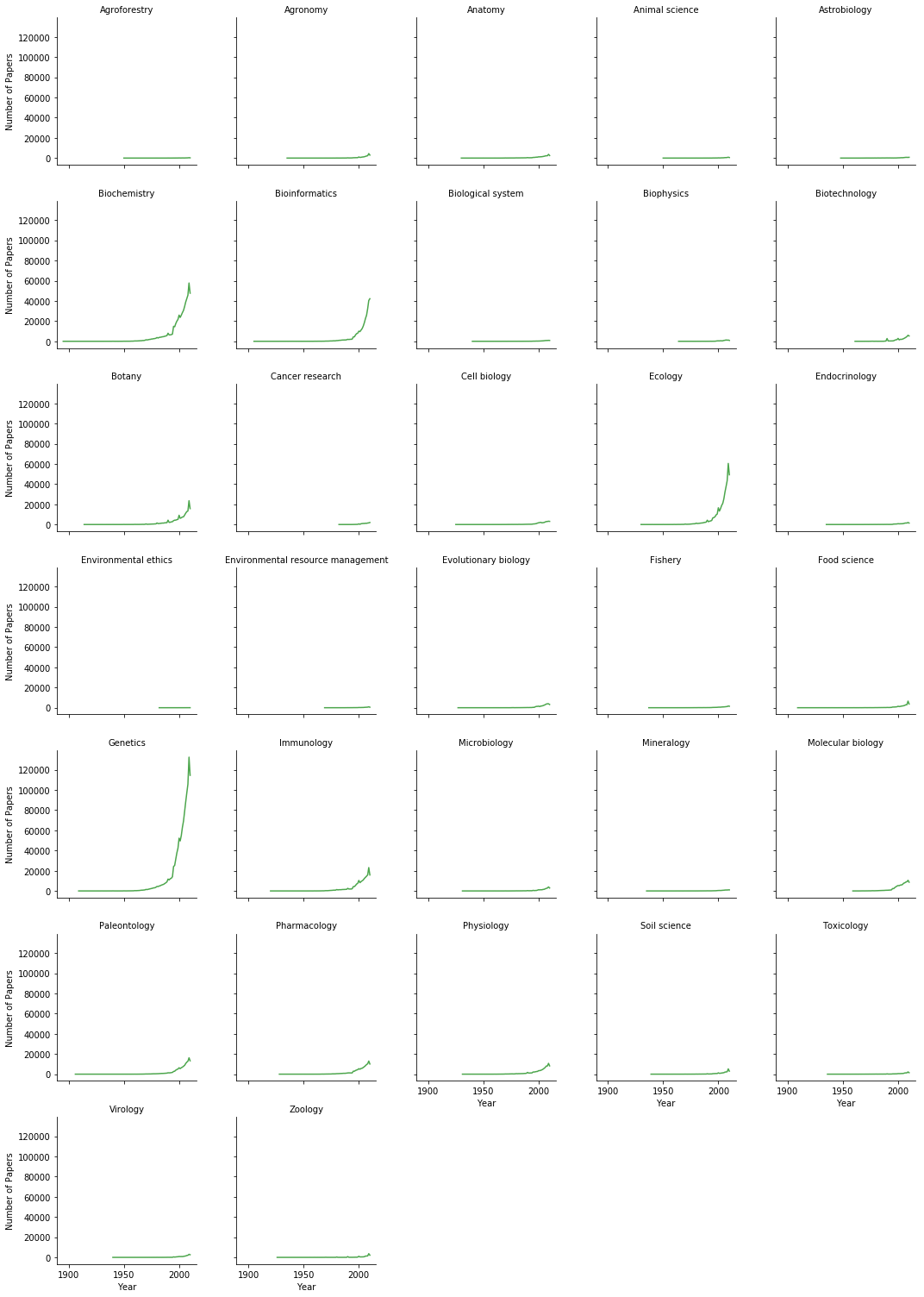}
\caption{\textbf{Biology L1-Subfields Number of Papers over Time.} We can observe a big variance in the number of papers over time in the various biology subfields.}
\label{fig:field_l1_num_papers_grid}
\end{figure*}

\begin{figure*}[ht]
\centering 
\includegraphics[width=0.8\linewidth]{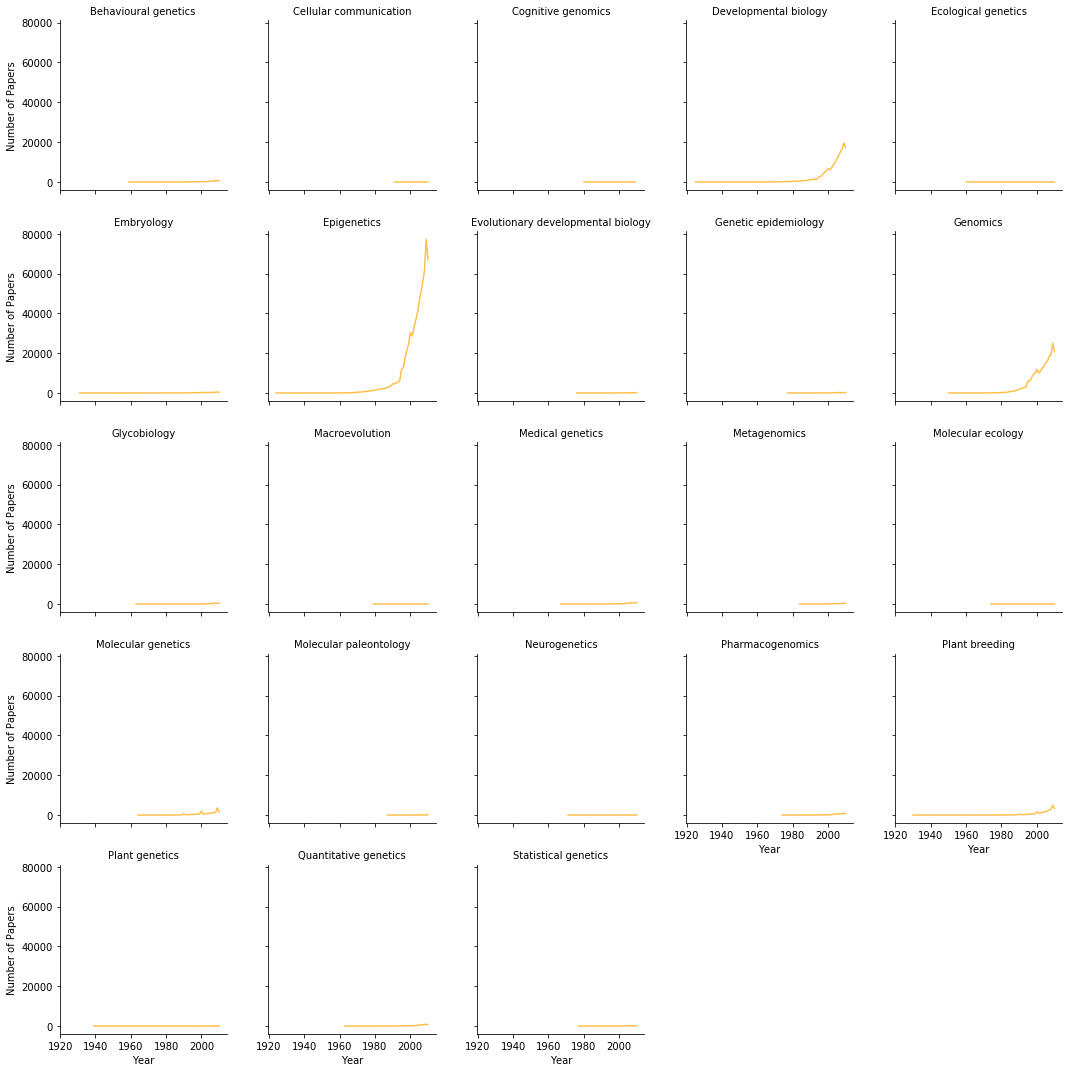}
\caption{\textbf{Genetics L2-Subfields Number of Papers over Time.} We can observe a big variance in the number of papers over time in the various genetics subfields.}
\label{fig:field_l2_num_papers_grid}

\end{figure*}

\begin{figure*}[ht]
\centering 
\includegraphics[width=0.8\linewidth]{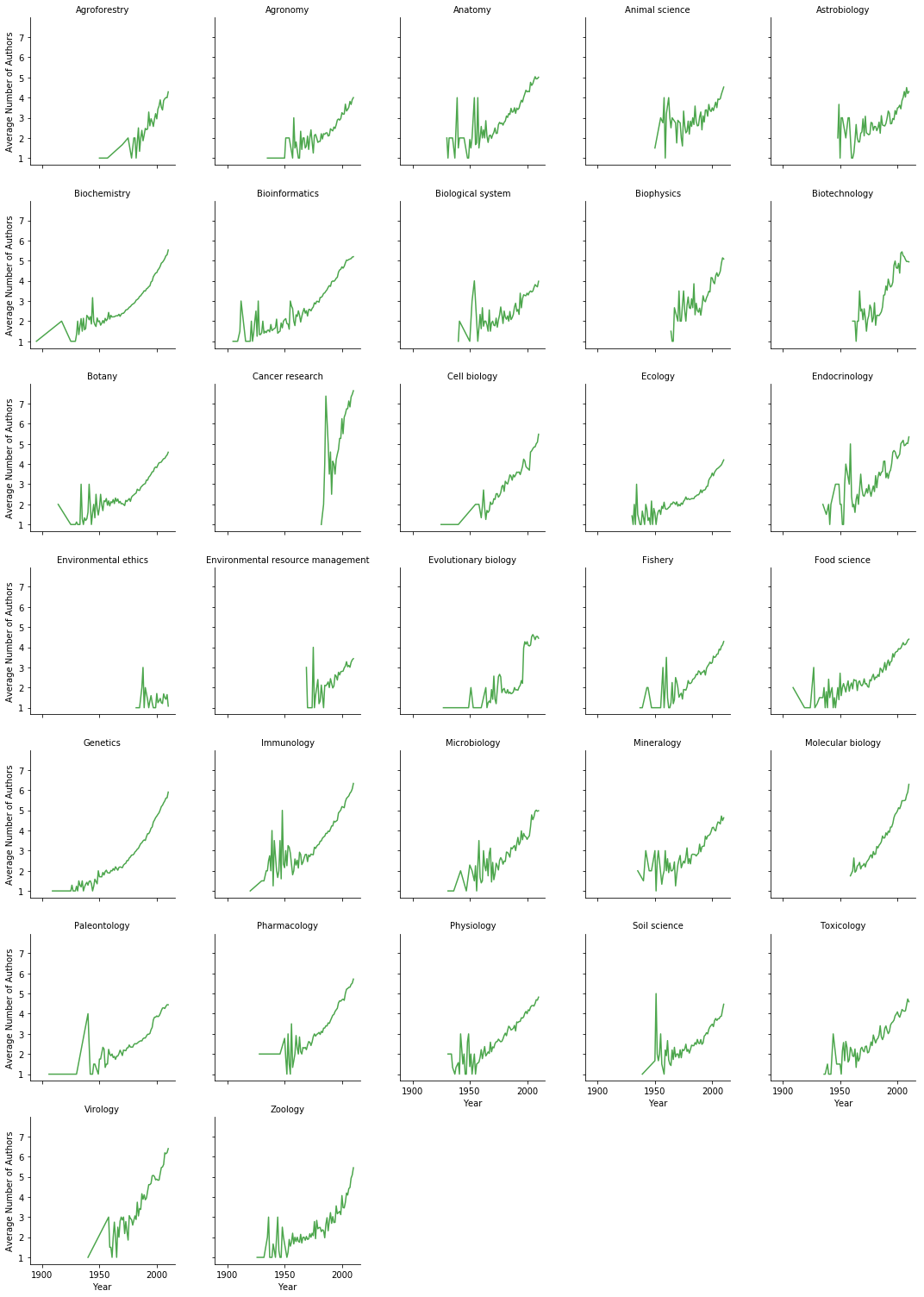}
\caption{ \textbf{Biology L1-Subfields Average Number of Authors over Time.} We can observe a variance in the average number of authors over time in the various biology subfields.}
\label{fig:field_l1_num_authors_grid}
\end{figure*}

\begin{figure*}[ht]
\centering 
\includegraphics[width=0.9\linewidth]{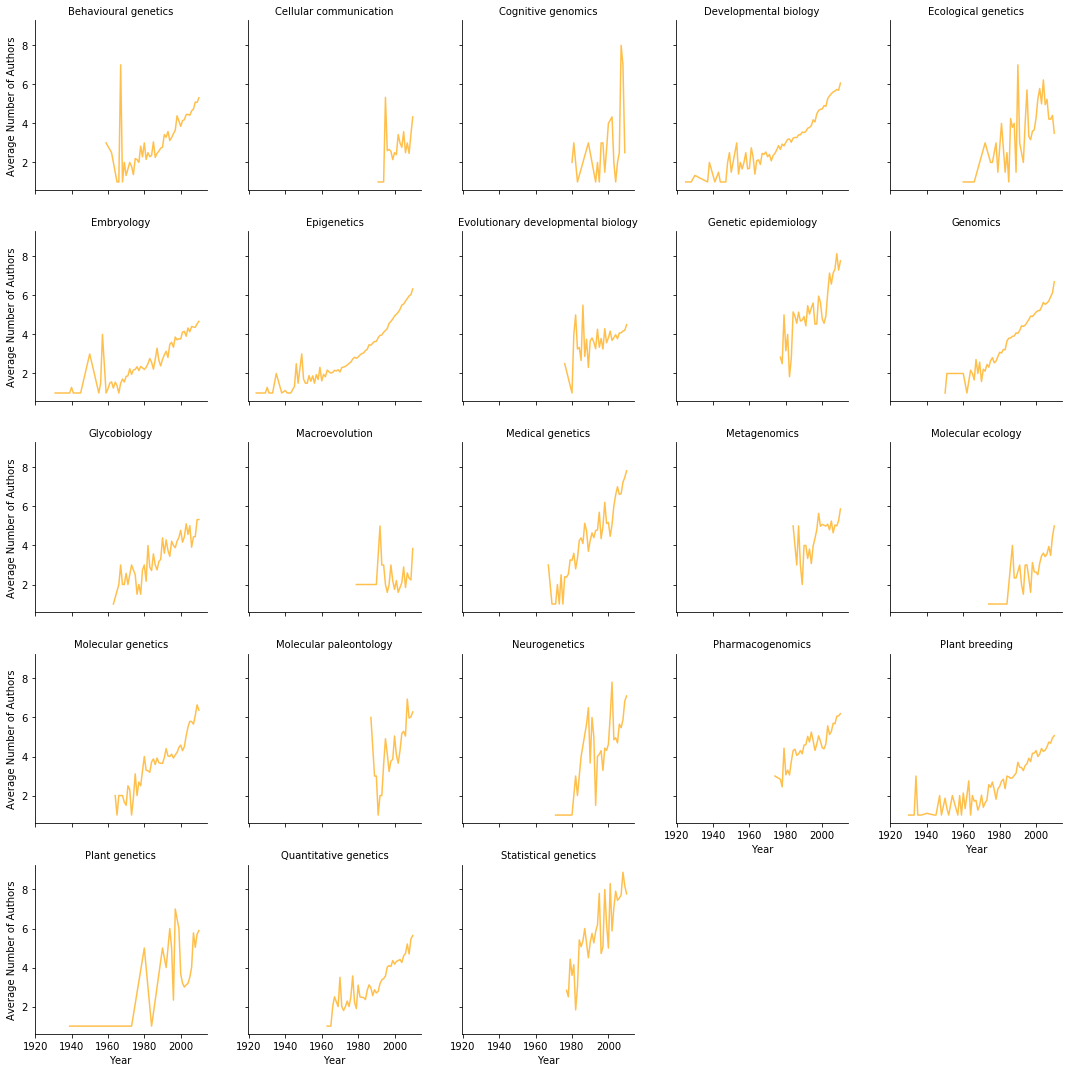}
\caption{\textbf{Genetics L3-Subfields Average Number of Authors over Time.} We can observe a significant variance in the average number of authors over time in the various genetics subfields.}
\label{fig:field_l2_num_authors_grid}

\end{figure*}

\begin{figure*}[ht]
\centering 
\includegraphics[width=0.8\linewidth]{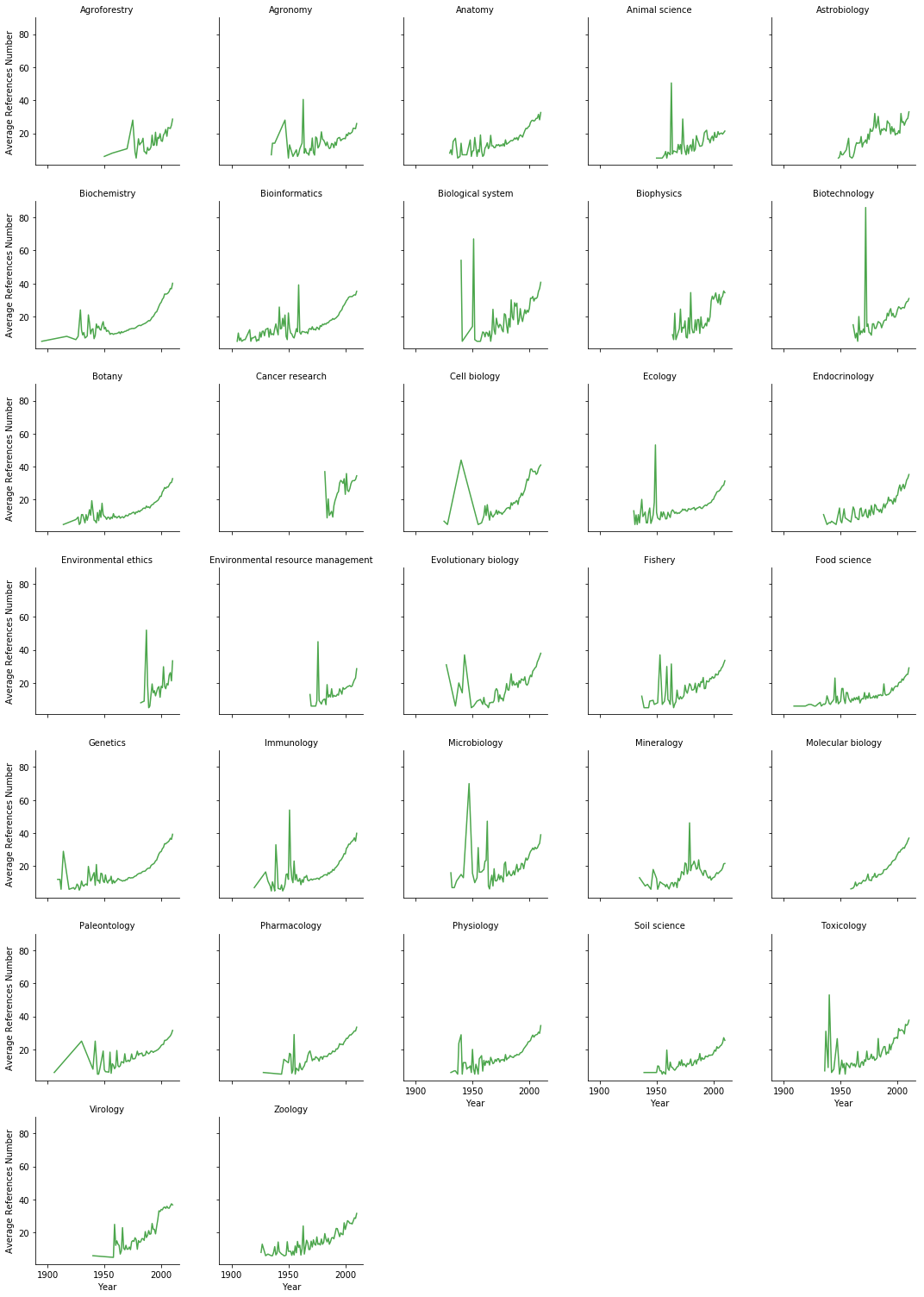}
\caption{ \textbf{Biology L1-Subfields Average Number of References over Time.} We can observe a variance in the average number of references over time in the various biology subfields.}
\label{fig:field_l1_num_ref_grid}
\end{figure*}

\begin{figure*}[ht]
\centering 
\includegraphics[width=0.9\linewidth]{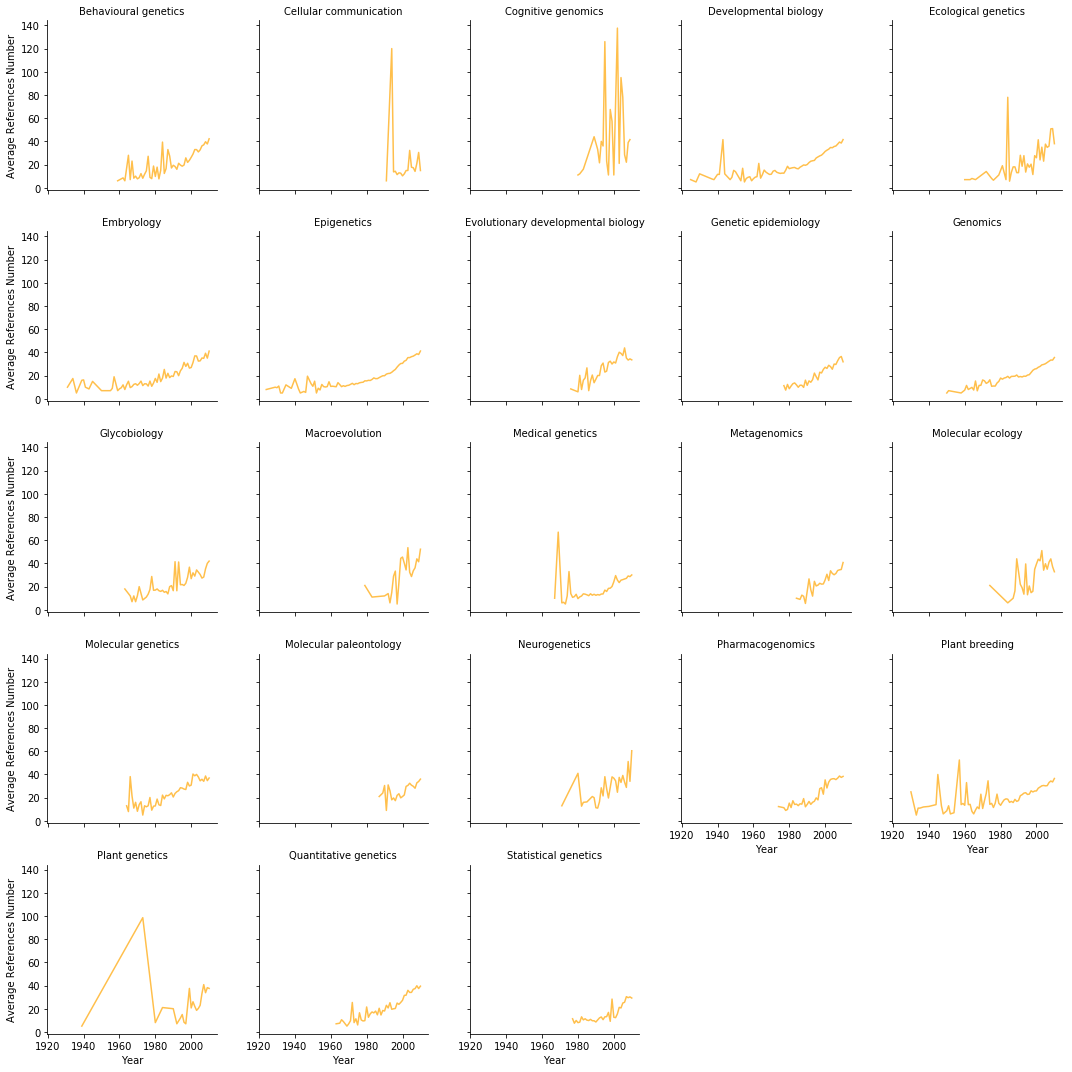}
\caption{\textbf{Genetics L2-Subfields Average Number of References over Time.} We can observe a significant variance in the average number of references over time in the various genetics subfields.}
\label{fig:field_l2_num_ref_grid}

\end{figure*}

\begin{figure*}[ht]
\centering 
\includegraphics[width=0.9\linewidth]{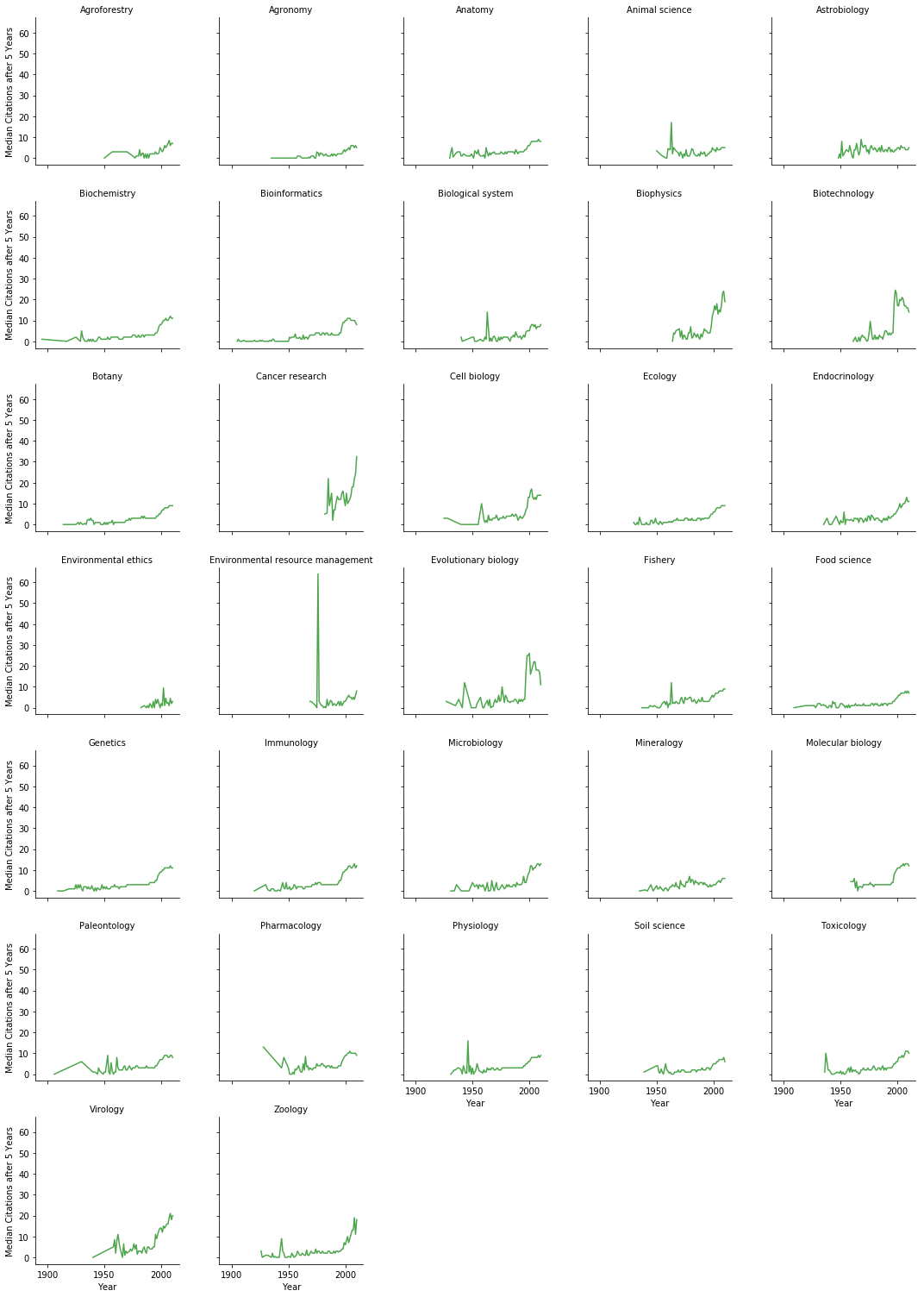}
\caption{ \textbf{Biology L1-Subfields Median Number of 5-Year Citations over Time.} We can observe a variance in the median number of citations over time in the various biology subfields. }
\label{fig:field_l1_num_citation_grid}
\end{figure*}

\begin{figure*}[ht]
\centering 
\includegraphics[width=0.9\linewidth]{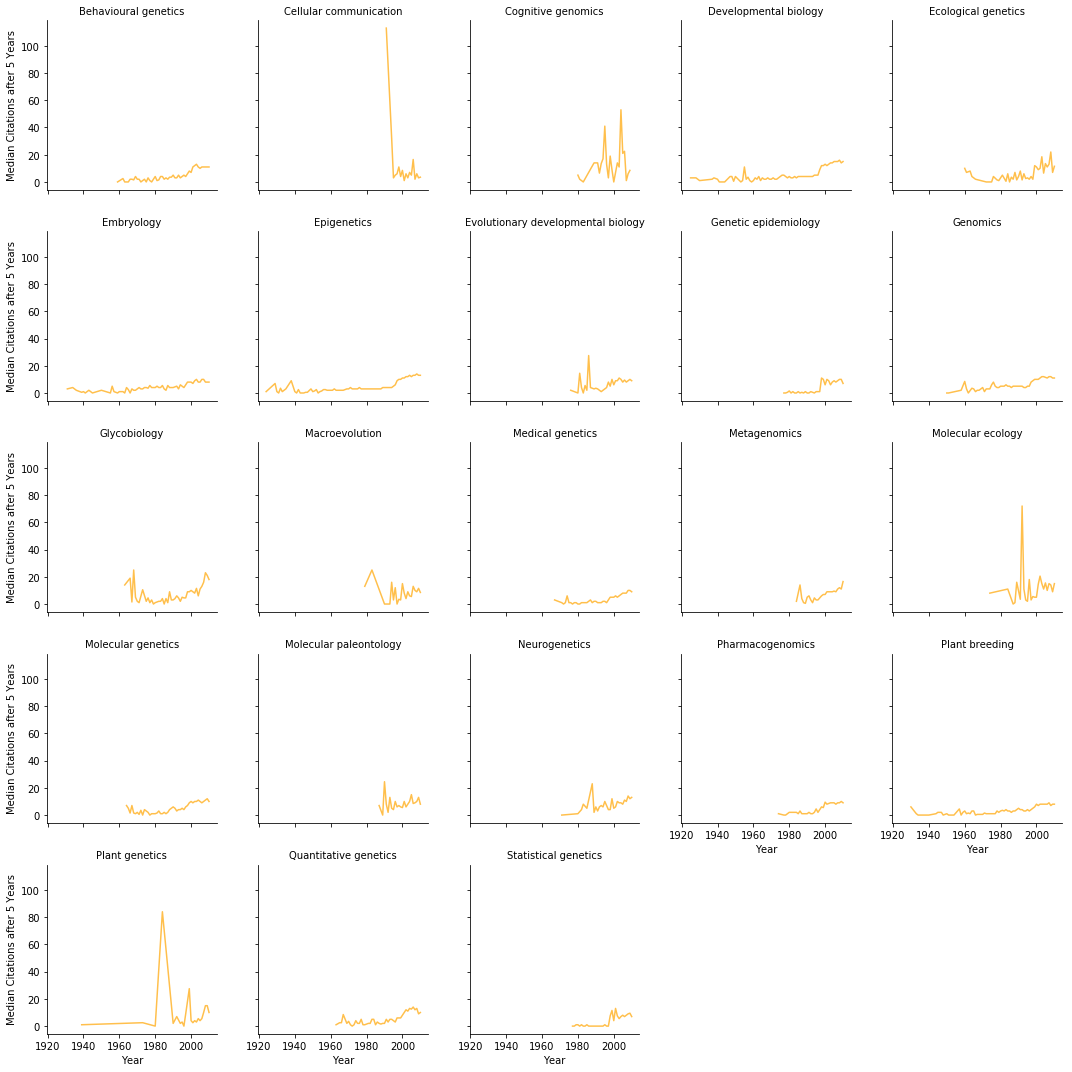}
\caption{\textbf{Genetics L2-Subfields Median Number of 5-Year Citations over Time.} We can observe a significant variance in the median number of citations over time in the various genetics subfields.}
\label{fig:field_l2_num_citation_grid}
\end{figure*}

\begin{figure*}[ht]
\centering 
\includegraphics[width=0.9\linewidth]{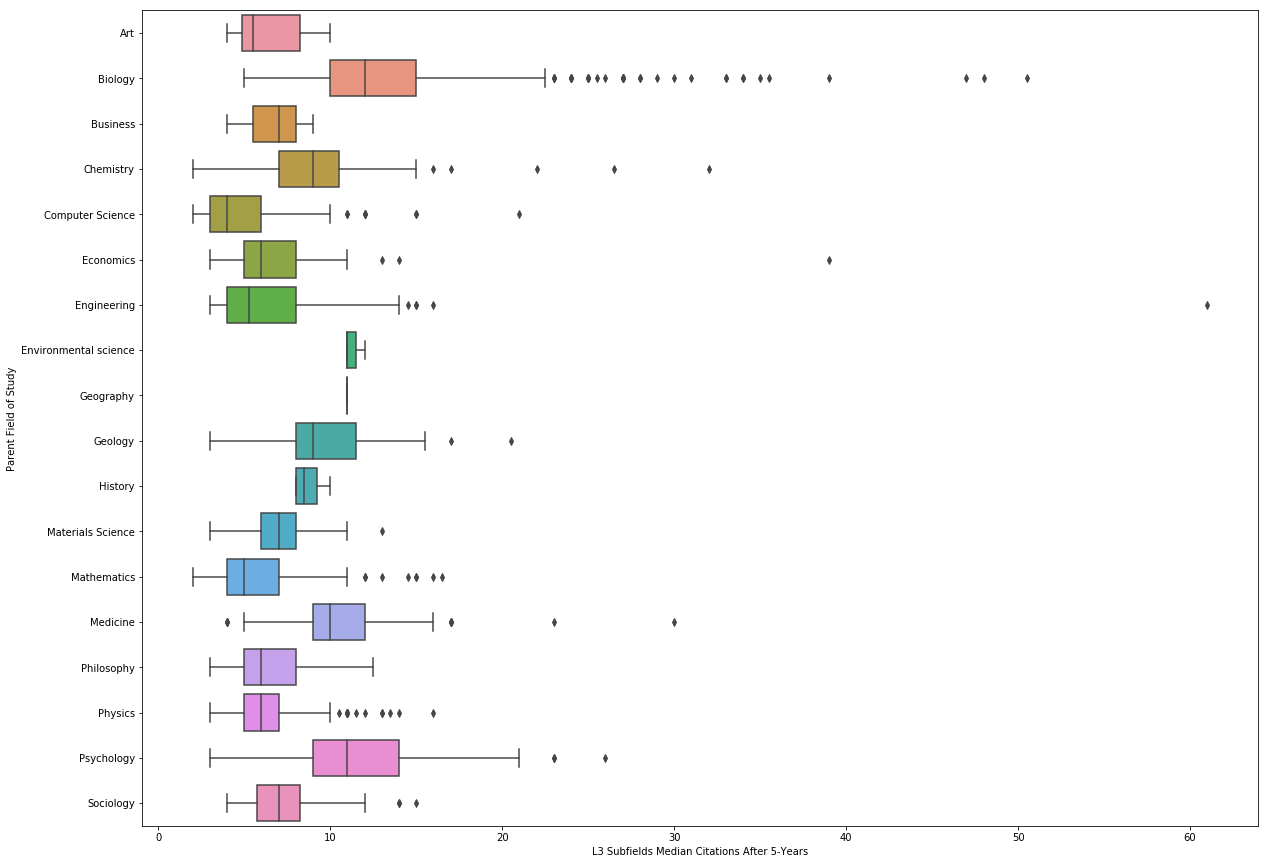}
\caption{\textbf{L3 Fields-of-Study Median 5-Year Citation Distributions by Parent Fields.} We can observe the high variance among the L3 fields-of-study median citation numbers. }
\label{fig:fields_l3_subs}
\end{figure*}

\begin{figure*}[ht]
\centering 
\includegraphics[width=0.9\linewidth]{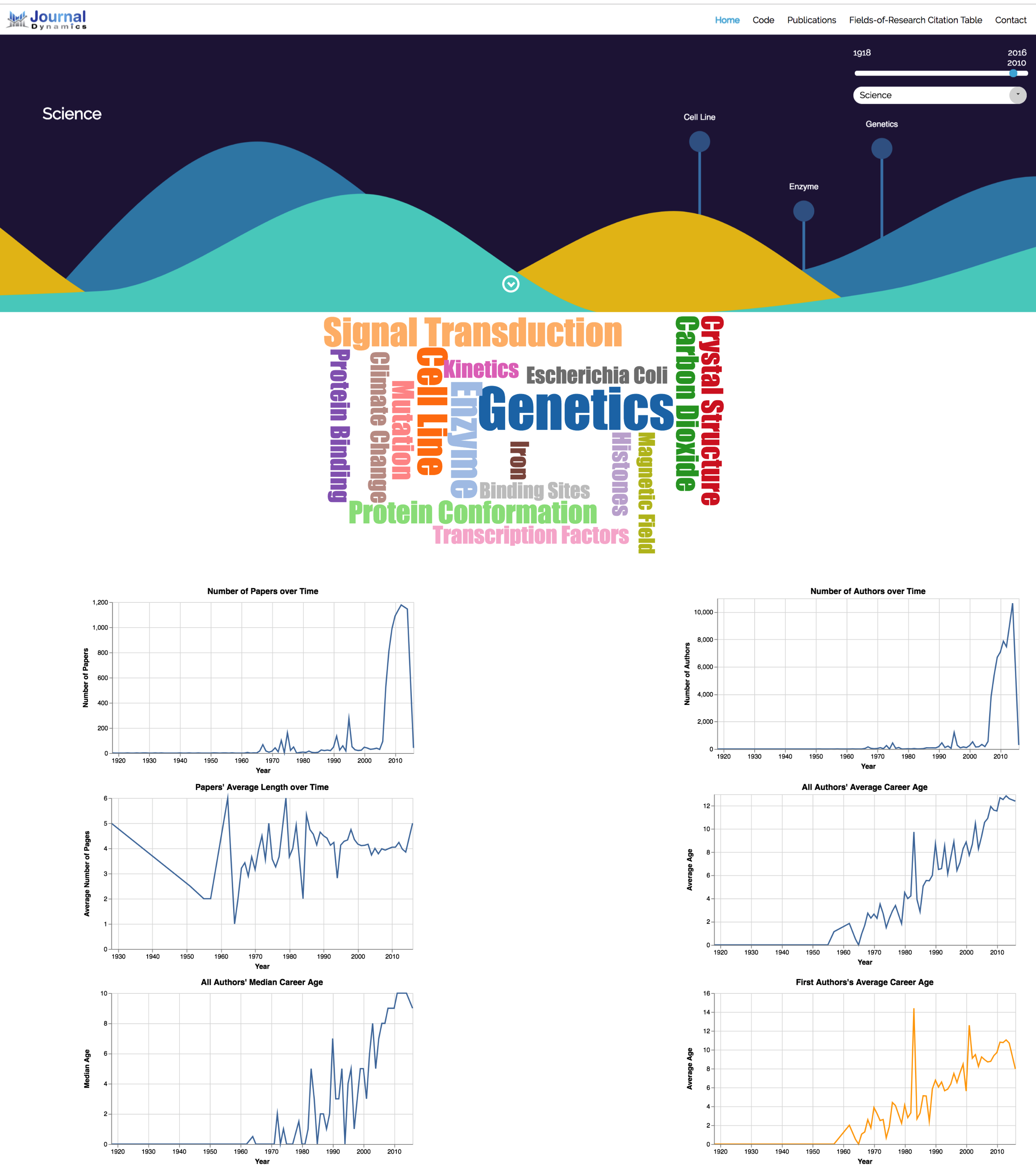}
\caption{\textbf{Interactive Website.} We have developed an \href{http://sciencedynamics.cs.washington.edu/}{interactive website} that makes it possible to view and interact directly with the study's data. }
\label{fig:website}
\end{figure*}

\end{document}